\documentclass[aps,prd,reprint,floatfix,amsmath,amssymb,nofootinbib,longbibliography]{revtex4-2}

\usepackage{graphicx}
\usepackage{xcolor}

\begin{document}
\raggedbottom

\title{Quasinormal Modes and Excitation Factors of a Parity-Symmetric Beyond-Horndeski Black Hole}

\author{Roman A. Konoplya}
\email{roman.konoplya@gmail.com}
\affiliation{Research Centre for Theoretical Physics and Astrophysics, Institute of Physics, Silesian University in Opava, Bezru\v{c}ovo n\'am\v{e}st\'i 13, CZ-74601 Opava, Czech Republic}

\author{Milena Skvortsova}
\email{milenas577@mail.ru}
\affiliation{People's Friendship University of Russia (RUDN University), 6 Miklukho-Maklaya Street, Moscow, 117198, Russia}

\begin{abstract}
We study scalar, electromagnetic, and Dirac perturbations of an asymptotically flat black hole in parity-symmetric beyond-Horndeski gravity. The quasinormal spectrum is computed with very-high-order WKB expansions and Pad\'e resummation, with Ces\`aro means used to diagnose internal convergence. The dominant  modes are independently tested by time-domain evolution with Prony extraction and by Chebyshev pseudospectral collocation; the latter also provides systematic frequency scans in both the metric-scale and physical ADM-mass normalizations. Excitation factors are obtained separately from Green-function residues by complex-contour integration. In the fixed metric-scale normalization, approaching Schwarzschild generally increases both the oscillation frequency and damping rate, whereas stronger deformation shifts the modes toward lower frequencies and weaker damping. In ADM-mass units, the mildly deformed solutions remain close to Schwarzschild, while the most deformed sector becomes enhanced and nonmonotonic because the asymptotic mass itself varies strongly. The independent methods agree very well, however the sixth-order WKB approximation used in [G. Antoniou, T. D. Pappas, P. Kanti, Phys. Rev. D
112, 084013 (2025)] is insufficient in certain regions of the parameter space: the fundamental-mode discrepancy reaches nine percent in the damping rate and the first-overtone discrepancy can be of order unity. The Schwarzschild excitation factors are recovered. In the strongly deformed regime, their magnitudes can differ from the Schwarzschild values by more than two orders of magnitude in the asymptotic tortoise-coordinate convention. We show that this large effect is predominantly due to the deformation-dependent displacement of the potential barrier in the tortoise coordinate: when its maximum is used as the origin, the excitation factors exhibit much more modest deviations.
\end{abstract}

\maketitle
\setlength{\parindent}{0pt}
\setlength{\parskip}{1\baselineskip}

\section{Introduction}

Quasinormal modes of black holes play a central role in modern gravitational physics because they characterize the intrinsic damped oscillations of perturbed compact objects and therefore encode direct information about the underlying spacetime geometry~\cite{Chandrasekhar:1983,Kokkotas:1999bd,Berti:2009kk,Konoplya:2011qq,Nollert:1999ji,Bolokhov:2025rng}. In astrophysical applications they govern the ringdown stage of black-hole mergers and provide one of the cleanest links between theory and observation, while from the theoretical side they offer a sensitive probe of stability, effective potentials, and deviations from general relativity. For this reason, the quasinormal spectrum is now widely used as a diagnostic of black-hole parameters and as a test of alternative theories of gravity.

The importance of the underlying gravitational theory is equally evident. Once one goes beyond general relativity, the background geometry, horizon structure, effective potentials, and therefore the quasinormal spectrum can all change in a theory-dependent way. Scalar-tensor and beyond-Horndeski models are especially interesting in this respect~\cite{Kobayashi:2019hrl,Gleyzes:2014dya}, because they provide consistent modifications of Einstein gravity while allowing nontrivial black-hole solutions whose ringdown properties can differ measurably from the Schwarzschild or Reissner--Nordstr\"om cases. Studying quasinormal modes in such theories is therefore a natural way to connect the mathematical structure of the field equations with potentially observable signatures of modified gravity.

In the present article we consider the black-hole background that appeared recently in analyses of greybody factors and quasinormal modes in parity-symmetric beyond-Horndeski gravity. The metric employed in Ref.~\cite{Antoniou:2025fpc} is the asymptotically flat exact solution first derived in Ref.~\cite{Bakopoulos:2022nqd}.
{Related beyond-Horndeski black holes have recently been studied from several complementary directions. Exact construction schemes and solutions carrying an independent primary scalar charge were developed in Refs.~\cite{Bakopoulos:2024primary,Baake:2023zsq}; their thermodynamics, shadows, strong-lensing and accretion-disk images, and gravitational perturbations were subsequently analyzed in Refs.~\cite{Erices:2025shadows,Fathi:2025lensing,Charmousis:2025xug,Bahamonde:2026iqb,Charmousis:2026bhe}. Test-field spectra, greybody factors, and Hawking radiation were calculated for another parity-symmetric primary-hair geometry in Ref.~\cite{Konoplya:2026cus}, while coupled gravitational and electromagnetic perturbations of charged vector-Horndeski black holes were treated in Ref.~\cite{Stashko:2026kja}. For the particular secondary-hair background considered here, the massive-scalar spectrum, absorption, quasi-resonances, and oscillatory late-time tails have also been investigated~\cite{Bolokhov:2026massive}. These studies concern different metrics or perturbation systems, but together they provide the relevant modified-gravity context for the present test-field analysis.}

{A recent complementary analysis of primary-hair black holes in shift-symmetric beyond-Horndeski gravity studied axial gravitational perturbations in both the physical- and effective-metric formulations, compared their time-domain evolutions and quasinormal spectra, and calculated the associated greybody factors and absorption cross sections~\cite{Antoniou:2026vrm}. Although the background and perturbation sector differ from those considered here, that work provides an important gravitational-sector counterpart to the present test-field calculation.}

{Our analysis is organized around complementary methods with distinct roles. Very-high-order WKB expansions, extending from tens of terms to one thousand in the most delicate case, together with Pad\'e resummation provide the broad scalar, electromagnetic, and Dirac spectra. Their internal reliability is assessed from the Ces\`aro behavior of the resummed sequence following Ref.~\cite{Konoplya:2026rjh}, rather than from a few neighboring orders~\cite{Bolokhov:2025egl,Bolokhov:2025aqy,Bolokhov:2025lnt,Skvortsova:2024msa,Malik:2026lfj,Lutfuoglu:2026uzy,Lutfuoglu:2026boa,Skvortsova:2026unq,Skvortsova:2026jtx,Dubinsky:2026wcv}. Time-domain evolution with Prony extraction and a factored Chebyshev pseudospectral calculation then provide independent checks of the frequencies, with the Chebyshev grid also allowing the spectrum to be compared in metric-scale and ADM-mass units. A multiprecision extension of the same boundary-value calculation follows the electromagnetic quadrupole through its first six overtones and tests the near-horizon overtone-sensitivity mechanism discussed in Ref.~\cite{Konoplya:2022pbc}. Finally, the validated pole locations are used in a complex-contour scattering calculation of the excitation factors. This unified strategy determines the deformation dependence of the spectrum, isolates the slow-convergence sector, and permits a quantitative reassessment of the earlier sixth-order WKB results of Antoniou, Pappas, and Kanti~\cite{Antoniou:2025fpc}.}

{Compared with Ref.~\cite{Antoniou:2025fpc}, we first demonstrate, using very-high-order WKB--Pad\'e calculations and independent time-domain and Chebyshev checks, that the sixth-order WKB approximation can produce large errors in part of the parameter space, reaching nine percent in the fundamental scalar-mode damping rate, while the first-overtone estimates can differ by order unity. We then present a systematic study of several lowest electromagnetic overtones and examine their sensitivity to the near-horizon geometry. Finally, we calculate excitation factors, thereby extending the analysis from the quasinormal frequencies to the residues of the scattering amplitudes. In particular, we distinguish the effect of the additive normalization of the tortoise coordinate from changes in the scattering properties of the potential itself by comparing a common asymptotic convention with one in which the potential maximum defines the origin.}

{The paper is organized as follows. Section~II describes the black-hole geometry, its asymptotic mass, and the admissible horizon structure, and Sec.~III derives the scalar, electromagnetic, and Dirac master equations. Sections~IV and V formulate the WKB--Pad\'e/Ces\`aro and time-domain/Prony calculations. Their spectral results and the comparison with sixth-order WKB are presented together in Sec.~VI. Section~VII develops the independent Chebyshev boundary-value calculation, reports the systematic scans in both mass normalizations, and tests the first six electromagnetic overtones of a representative deformation. These validated bosonic poles are used in Sec.~VIII to calculate the excitation factors, after which Sec.~IX summarizes the conclusions.}

\section{Parity-symmetric beyond-Horndeski black hole}

We begin from the shift-symmetric, parity-symmetric sector of beyond-Horndeski theory in four dimensions, for which
the action, the restricted Lagrangian densities, and the notation follow Ref.~\cite{Bakopoulos:2022nqd}.
\begin{equation}
S = \int d^4x\,\sqrt{-g}\,\bigl(\mathcal{L}_2 + \mathcal{L}_4 + \mathcal{L}^{\rm bH}_4\bigr),
\end{equation}
with
\begin{align}
\mathcal{L}_2 &= G_2(X), \\
\mathcal{L}_4 &= G_4(X)R + G_{4X}(X)\Bigl[(\Box\phi)^2 - \phi_{\mu\nu}\phi^{\mu\nu}\Bigr], \\
\mathcal{L}^{\rm bH}_4 &= F_4(X)\,\varepsilon^{\mu\nu\rho}{}_{\sigma}\,\varepsilon^{\mu'\nu'\rho'\sigma}\,\phi_{\mu}\phi_{\mu'}\phi_{\nu\nu'}\phi_{\rho\rho'}.
\end{align}
The three pieces entering the action have a clear interpretation. The term $\mathcal{L}_2$ contains the purely kinetic part of the scalar sector through the arbitrary function $G_2(X)$. The term $\mathcal{L}_4$ is the quartic Horndeski contribution: its first part, proportional to $G_4(X)$, gives a nonminimal coupling between the scalar field and the Ricci scalar, while the second part, proportional to $G_{4X}(X)$, contains the specific combination of second derivatives that preserves the degeneracy structure of the theory. Finally, $\mathcal{L}^{\rm bH}_4$ is the quartic beyond-Horndeski correction governed by $F_4(X)$; it is built from contractions of the Levi-Civita tensors with first and second derivatives of the scalar field and represents the genuinely beyond-Horndeski sector.

Here
\begin{equation}
X \equiv -\frac{1}{2}\nabla_{\mu}\phi\nabla^{\mu}\phi, \qquad
\phi_{\mu} \equiv \nabla_{\mu}\phi, \qquad
\phi_{\mu\nu} \equiv \nabla_{\mu}\nabla_{\nu}\phi,
\end{equation}
so that $X$ is the standard kinetic density of the scalar field, $\phi_{\mu}$ denotes its first covariant derivative, and $\phi_{\mu\nu}$ its second covariant derivative, whereas $G_{4X}$ stands for $dG_4/dX$.
In the full shift-symmetric beyond-Horndeski/DHOST construction one may also introduce the additional free functions $G_3(X)$, $G_5(X)$, and $F_5(X)$; however, in the parity-symmetric subsector considered here these odd-parity pieces are consistently absent, so one sets $G_3 = G_5 = F_5 = 0$~\cite{Bakopoulos:2022nqd}.

For a static, spherically symmetric configuration,
\begin{equation}
ds^2 = -h(r)dt^2 + \frac{dr^2}{f(r)} + r^2(d\theta^2 + \sin^2\theta\,d\varphi^2),
\quad \phi = \phi(r),
\end{equation}
the field equations simplify considerably after introducing the auxiliary function
\begin{equation}
Z(X) \equiv 2XG_{4X} - G_4 + 4X^2F_4.
\end{equation}
On the integrable branch used in Ref.~\cite{Bakopoulos:2022nqd}, one takes $Z=\gamma$ with $\gamma$ constant. The reduced equations then imply
\begin{equation}
Z^2f = \gamma^2 h,
\end{equation}

and hence $f=h$ on the constant-$Z$ branch; the asymptotically flat solution selected below further requires $\gamma=-1$. A solvable family follows from the compatibility construction
\begin{align}
G_2Z&=\epsilon\mathcal{G}-\alpha X+C,\\
2G_4Z&=\delta\mathcal{G}-\beta X+D.
\end{align}
\begin{equation}
\mathcal{G}_X=2\mu X+\zeta.
\end{equation}
Here $\mathcal{G}(X)$ is an auxiliary function introduced by the compatibility conditions above, while $C$ and $D$ are integration constants whose combinations are fixed below by the constant terms in $G_2$ and $G_4$. The quantities $\beta$, $\mu$, $\delta$, $\epsilon$, $\zeta$, and $\alpha$ are constant parameters of the underlying theory in the notation of Ref.~\cite{Bakopoulos:2022nqd}, while $\Lambda$ plays the role of the cosmological constant term and $\gamma$ labels the constant value of the branch function $Z(X)$. Once these constants are fixed, the metric and scalar profile are determined by the exact solution below, which yields
\begin{align}
G_2(X) &= \frac{\epsilon\mu}{\gamma}X^2 + \frac{\epsilon\zeta - \alpha}{\gamma}X - 2\Lambda, \\
G_4(X) &= \frac{\delta\mu}{2\gamma}X^2 + \frac{\delta\zeta - \beta}{2\gamma}X + 1, \\
F_4(X) &= \frac{\gamma+1}{4X^2} + \frac{\beta - \delta\zeta}{8\gamma X} - \frac{3\delta\mu}{8\gamma},
\end{align}
and the scalar kinetic density
\begin{equation}
X(r) = \frac{(\alpha - \epsilon\zeta)r^2 + \beta - \delta\zeta}{2\mu(\epsilon r^2 + \delta)}.
\end{equation}

The asymptotically flat branch relevant for QNM applications is obtained by imposing
\begin{equation}
\alpha = \epsilon\zeta, \qquad \Lambda = 0, \qquad \gamma=-1.
\end{equation}
With these choices the action functions become
\begin{align}
G_2(X) &= -\epsilon\mu X^2, \\
G_4(X) &= -\frac{\delta\mu}{2}X^2 + \frac{\beta - \delta\zeta}{2}X + 1, \\
F_4(X) &= \frac{\delta\zeta - \beta}{8X} + \frac{3\delta\mu}{8},
\end{align}
while the exact line element, corresponding to the asymptotically flat branch of the solution derived in Ref.~\cite{Bakopoulos:2022nqd}, reduces to
\begin{equation}\label{eq:line-element}
ds^2 = -A(r)dt^2 + \frac{dr^2}{A(r)} + r^2(d\theta^2 + \sin^2\theta\,d\varphi^2),
\end{equation}
with
\begin{equation}\label{eq:metric-function}
A(r) = 1 + p_2\frac{\arctan(p_1 r)}{p_1 r} - \frac{2M}{r}.
\end{equation}
The two parameters entering the metric are
\begin{equation}
p_1 = \sqrt{\frac{\epsilon}{\delta}} > 0, \qquad
p_2 = \frac{(\beta - \delta\zeta)^2}{8\delta\mu},
\end{equation}
Thus $p_1$ has dimensions of inverse length, whereas $p_2$ is dimensionless; numerical values of $p_1$ below are quoted in units with $M=1$.
The scalar profile obeys
\begin{equation}
\phi'(r)^2 = -\frac{\beta - \delta\zeta}{\mu(\epsilon r^2 + \delta)}\,\frac{1}{A(r)}.
\end{equation}
The reality of the scalar therefore requires $(\beta-\delta\zeta)$ and $\delta\mu$ to have opposite signs. In the phenomenologically most interesting branch one has $p_2<0$, which also guarantees a Reissner--Nordstr\"om-like asymptotic expansion.

\subsection{Main properties of the metric}

Several basic features of the geometry follow directly from Eq.~\eqref{eq:metric-function}.

First, the Schwarzschild solution is recovered smoothly in the limit $p_2 \to 0$:
\begin{equation}
A(r) \longrightarrow 1 - \frac{2M}{r}.
\end{equation}
Hence $p_2$ measures the deviation from Schwarzschild, while $p_1^{-1}$ sets the scale at which the arctangent deformation becomes important. One also recovers the Schwarzschild limit when $p_1\to\infty$ at fixed $p_2$ and finite $r$, because $\arctan(p_1 r)/(p_1 r)\to 0$ and therefore again $A(r)\to 1-2M/r$.

Second, at large radius one finds
\begin{equation}
\frac{\arctan(p_1 r)}{p_1 r} = \frac{\pi}{2p_1 r} - \frac{1}{p_1^2r^2} + \mathcal{O}(r^{-4}),
\end{equation}
so that
\begin{equation}
A(r) = 1 - \frac{2M_{\rm ADM}}{r} + \frac{Q^2}{r^2} + \mathcal{O}(r^{-4}),
\end{equation}
with the effective ADM mass and tidal-charge parameter given by
\begin{equation}
M_{\rm ADM} = M - \frac{\pi p_2}{4p_1},
\qquad
Q^2 = -\frac{p_2}{p_1^2}.
\end{equation}
This large-radius expansion makes clear that the parameter $M$ entering the exact metric is not identical to the asymptotic mass measured at infinity. Rather, the physical mass seen in the $1/r$ falloff is $M_{\rm ADM}$, which differs from $M$ by the deformation-dependent shift $-\pi p_2/(4p_1)$. Only in the Schwarzschild limit $p_2 \to 0$ does one recover $M_{\rm ADM}=M$.
{Accordingly, the dimensionless frequency appropriate for comparing different members of the family by their mass measured at infinity is
\begin{equation}\label{eq:adm-frequency}
\widehat{\omega}_{\rm ADM}\equiv M_{\rm ADM}\omega.
\end{equation}
A scan performed at fixed $M$ does not keep the physical mass fixed, because $M_{\rm ADM}/M=1-\pi p_2/(4p_1M)$ changes with the deformation. We therefore retain the original fixed-$M$ frequencies for comparison with the WKB and time-domain calculations, but report the Chebyshev results also as $M_{\rm ADM}\omega$.}
Therefore, when $p_2<0$ the geometry approaches a Reissner--Nordstr\"om-like black hole with positive effective charge $Q^2>0$.

Third, near the centre,
\begin{equation}
\frac{\arctan(p_1 r)}{p_1 r} = 1 - \frac{p_1^2r^2}{3} + \mathcal{O}(r^4),
\end{equation}
which implies
\begin{equation}
A(r) = 1 + p_2 - \frac{2M}{r} - \frac{p_2p_1^2}{3}r^2 + \mathcal{O}(r^4).
\end{equation}
For $M\neq 0$ the $-2M/r$ term dominates and the origin remains singular, so the beyond-Horndeski correction does not regularize the central curvature singularity. The scalar hair is therefore secondary: the geometry is still controlled by the mass parameter $M$, while $p_1$ and $p_2$ are fixed by couplings in the underlying action rather than by an independent conserved scalar charge.

\subsection{Horizon structure and allowed parameter ranges}

The event horizon radius $r_h$ is obtained from $A(r_h)=0$. Introducing the dimensionless variables
\begin{equation}
x \equiv p_1 r, \qquad m \equiv p_1 M,
\end{equation}
the equation for a horizon becomes
\begin{equation}\label{eq:horizon-equation}
x + p_2\arctan x - 2m = 0, \qquad x>0.
\end{equation}
Its derivative is
\begin{equation}
\frac{d}{dx}\Bigl(x + p_2\arctan x\Bigr) = 1 + \frac{p_2}{1+x^2}.
\end{equation}
This expression is enough to classify the causal structure.

If $p_2\geq -1$, the function in Eq.~\eqref{eq:horizon-equation} is monotonic increasing for all $x>0$. In that case:
\begin{itemize}
    \item for $M>0$, there is exactly one event horizon;
    \item for $M\leq 0$, there is no finite positive-radius horizon.
\end{itemize}

If $p_2<-1$, the left-hand side of Eq.~\eqref{eq:horizon-equation} develops a single minimum at
\begin{equation}
x_0 = \sqrt{-1-p_2}.
\end{equation}
The value of the left-hand side at this minimum is
\begin{equation}
\mathcal{F}_{\rm min} = x_0 + p_2\arctan x_0 - 2m.
\end{equation}
The horizon structure is then:
\begin{itemize}
    \item for $M>0$, there is still a single event horizon;
    \item for $M=0$, a horizon exists if and only if $p_2<-1$;
    \item for $M<0$, horizons exist only when $\mathcal{F}_{\rm min}\leq 0$, namely when
    \begin{equation}
    \begin{split}
    m \geq m_{\rm ext}(p_2) \equiv {}& \frac{1}{2}\Bigl[\sqrt{-1-p_2} \\
    &+ p_2\arctan\bigl(\sqrt{-1-p_2}\bigr)\Bigr].
    \end{split}
    \end{equation}
    Equality gives an extremal black hole with a degenerate horizon, while $m_{\rm ext}(p_2)<m<0$ yields two horizons and $m<m_{\rm ext}(p_2)$ gives a naked singularity.
\end{itemize}

For the special case $M=0$, the condition $p_2<-1$ is equivalent to
\begin{equation}
-\frac{(\beta-\delta\zeta)^2}{8} < \delta\mu < 0,
\end{equation}
which is precisely the parameter window emphasized in the original derivation of the massless scalar-hairy black hole~\cite{Bakopoulos:2022nqd}. In practice, for QNM calculations one usually focuses on the simpler positive-mass branch $M>0$ with $p_1>0$ and $p_2<0$, because it always possesses an event horizon and interpolates smoothly between the Schwarzschild limit and a Reissner--Nordstr\"om-like asymptotic regime.

The metric~\eqref{eq:metric-function} will be the background geometry for the perturbation equations studied later. The two deformation parameters $p_1$ and $p_2$ capture, respectively, the radial scale of the beyond-Horndeski modification and the strength/sign of the deviation from Schwarzschild, while the horizon condition above provides the parameter domain in which the QNM problem is well posed.

\section{Perturbations of fields of various spin}

We now consider fields propagating on the fixed black-hole background introduced in Sec.~II, namely the line element~\eqref{eq:line-element} with metric function~\eqref{eq:metric-function}.
Since $g_{tt}g_{rr}=-1$, all perturbation equations can be reduced to a Schr\"odinger-like form once the tortoise coordinate is introduced,
\begin{equation}
\frac{dr_*}{dr} = \frac{1}{A(r)}.
\end{equation}
For quasinormal modes one imposes purely ingoing waves at the event horizon and purely outgoing waves at spatial infinity,
\begin{equation}
\begin{aligned}
\Psi(r) &\sim e^{-i\omega r_*} \quad (r\to r_h), \\
\Psi(r) &\sim e^{+i\omega r_*} \quad (r\to \infty).
\end{aligned}
\end{equation}

\subsection{Scalar and electromagnetic perturbations}

For massless bosonic fields of spin $s=0,1$, the perturbation equations can be written in a unified way. Following the standard separation (see~\cite{Carter:1968ks,Konoplya:2018arm} and references therein), one expands the field as
\begin{equation}
\Phi_s(t,r,\theta,\varphi) = e^{-i\omega t}\,Y_{\ell m}(\theta,\varphi)\,\frac{\Psi_s(r)}{r},
\end{equation}
for the scalar sector, while in the electromagnetic case one decomposes the Maxwell potential into vector spherical harmonics. In both cases the radial function satisfies
\begin{equation}\label{eq:bosonic-master}
\frac{d^2\Psi_s}{dr_*^2} + \bigl[\omega^2 - V_s(r)\bigr]\Psi_s = 0,
\end{equation}
with effective potential
\begin{equation}\label{eq:bosonic-potential}
V_s(r) = A(r)\left[\frac{\ell(\ell+1)}{r^2} + (1-s^2)\frac{A'(r)}{r}\right].
\end{equation}
Therefore the explicit scalar and electromagnetic potentials are
\begin{align}
V_{\rm sc}(r) &= A(r)\left[\frac{\ell(\ell+1)}{r^2} + \frac{A'(r)}{r}\right], \qquad \ell=0,1,2,\dots, \\
V_{\rm em}(r) &= A(r)\frac{\ell(\ell+1)}{r^2}, \qquad \ell=1,2,3,\dots.
\end{align}

\subsection{Massless Dirac perturbations}

For a massless spinor, the starting point is the curved-space Dirac equation
\begin{equation}
\gamma^a e_a{}^{\mu}\bigl(\partial_{\mu}+\Gamma_{\mu}\bigr)\Psi_{1/2} = 0,
\end{equation}
where $e_a{}^{\mu}$ is a tetrad and $\Gamma_{\mu}$ is the spin connection. After the standard rescaling by $A^{-1/4}$ and separation in spinor spherical harmonics~\cite{Chandrasekhar:1983}, the radial sector is described by two first-order functions $F$ and $G$ obeying
\begin{align}
\left(\frac{d}{dr_*} - i\omega\right)F &= W(r)\,G, \\
\left(\frac{d}{dr_*} + i\omega\right)G &= W(r)\,F,
\end{align}
with superpotential
\begin{equation}
W(r) = \frac{\lambda\sqrt{A(r)}}{r},
\qquad \lambda = 1,2,3,\dots.
\end{equation}
Here $\lambda=\ell+1/2=1,2,3,\dots$ is the positive integer separation constant, with the spinor angular quantum number $\ell=1/2,3/2,5/2,\dots$.
Introducing the combinations $Z_+=F+G$ and $Z_-=F-G$, one obtains two decoupled Schr\"odinger-like equations,
\begin{equation}
\begin{split}
\frac{d^2Z_+}{dr_*^2} + \bigl[\omega^2 - V_D^{+}(r)\bigr]Z_+ &= 0, \\
\frac{d^2Z_-}{dr_*^2} + \bigl[\omega^2 - V_D^{-}(r)\bigr]Z_- &= 0.
\end{split}
\end{equation}
with supersymmetric partner potentials
\begin{equation}\label{eq:dirac-potentials}
\begin{split}
V_D^{\pm}(r) &= W(r)^2 \pm \frac{dW}{dr_*} \\
&= \frac{\lambda^2 A(r)}{r^2}
\pm \lambda\sqrt{A(r)}\left[\frac{A'(r)}{2r} - \frac{A(r)}{r^2}\right].
\end{split}
\end{equation}
The two potentials $V_D^{\pm}$ are isospectral, so either one may be used in the QNM calculation.

Collecting the three cases, scalar, electromagnetic, and massless Dirac perturbations all reduce to the one-dimensional master equation
\begin{equation}\label{eq:generic-master}
\frac{d^2\Psi_i}{dr_*^2} + \bigl[\omega^2 - V_i(r)\bigr]\Psi_i = 0,
\end{equation}
with the corresponding potentials given by Eqs.~\eqref{eq:bosonic-potential} and~\eqref{eq:dirac-potentials}. {This common form allows the frequency calculation to be organized consistently: WKB--Pad\'e is applied to all three sectors, time-domain evolution provides direct waveform checks, Chebyshev collocation supplies an independent  boundary-value calculation, and the Jost solutions determine the excitation factors.}

\begin{figure*}[t]
\centering
\includegraphics[width=0.98\textwidth]{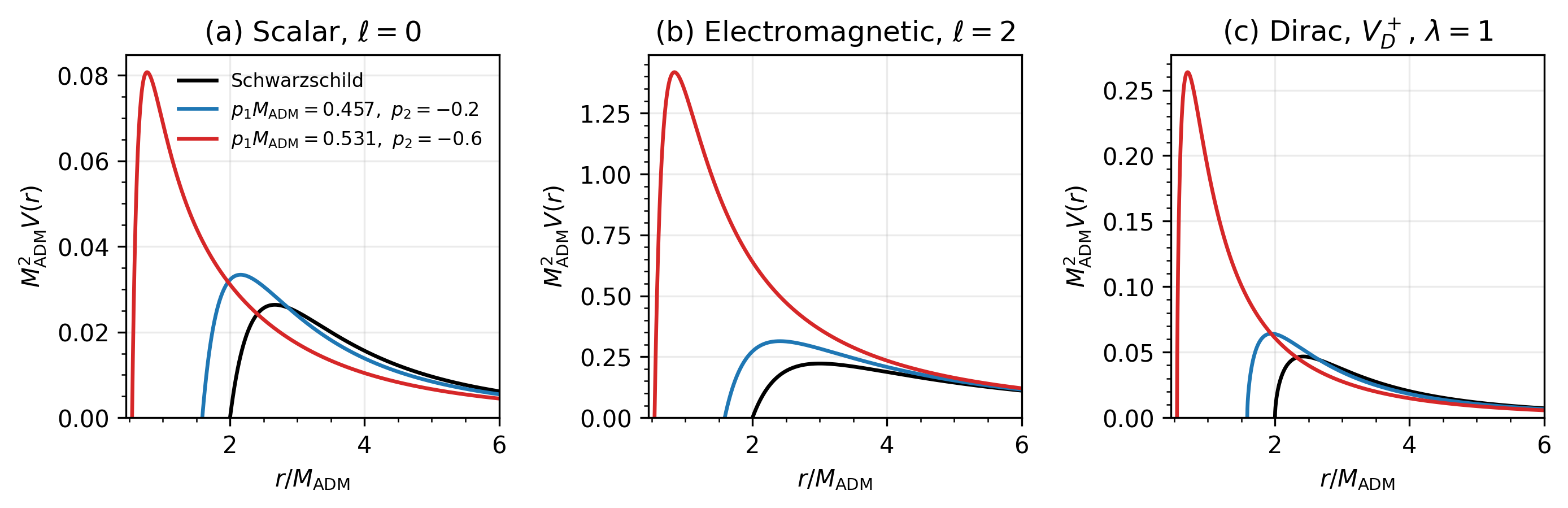}
\caption{Representative effective potentials in unit-ADM-mass variables. The axes are $r/M_{\rm ADM}$ and $M_{\rm ADM}^2V$, so each curve describes a black hole rescaled to $M_{\rm ADM}=1$. Panels (a), (b), and (c) show the scalar monopole, electromagnetic quadrupole, and lowest Dirac $V_D^+$ potential, respectively. The black curves are Schwarzschild. The blue and red curves have $(p_1M_{\rm ADM},p_2)=(0.457,-0.2)$ and $(0.531,-0.6)$; before rescaling, these correspond to the $M=1$ choices $(p_1,p_2)=(0.3,-0.2)$ and $(0.06,-0.6)$.}
\label{fig:effective-potentials}
\end{figure*}

{Representative profiles are displayed in Fig.~\ref{fig:effective-potentials} after rescaling every background to unit ADM mass. For all three perturbing fields, the potential shown vanishes at the event horizon and at spatial infinity and forms a single positive barrier in the exterior region. The moderately deformed barriers are visibly higher and lie inward of their Schwarzschild counterparts, but remain intermediate between the Schwarzschild and strongly deformed profiles. The latter exhibit a much larger enhancement at smaller $r/M_{\rm ADM}$, in accord with the ADM-normalized frequencies found in Sec.~VII. For the Dirac field we plot $V_D^+$, since its partner $V_D^-$ is isospectral.}

\section{Higher-order WKB method and Pad\'e resummation}

{
We compute the quasinormal frequencies from the unified master equation~\eqref{eq:generic-master} using the higher-order WKB expansion and its Pad\'e improvement~\cite{Iyer:1986np,Konoplya:2003ii,Matyjasek:2017psv,Matyjasek:2019eeu,Konoplya:2026wkb}. For a single-barrier potential with maximum at $r_{*0}$, the order-$N$ WKB condition can be written in the standard compact form
\begin{equation}\label{eq:wkb-quantization}
i\frac{\omega^2-V_0}{\sqrt{-2V_2}}
-\sum_{j=2}^{N}\Lambda_j
=n+\frac{1}{2},
\end{equation}
where $V_k=d^kV/dr_*^k|_{r_{*0}}$ and the correction terms $\Lambda_j$ are constructed from the derivatives through $V_{2j}$. Their explicit expressions are standard and are not reproduced here.

The resulting formal high-order series for $\omega^2$ is asymptotic~\cite{HatsudaBorel} and is resummed directly with balanced diagonal or near-diagonal Pad\'e approximants,
\begin{equation}\label{eq:pade-def}
\begin{aligned}
P_{L/M}(\epsilon)
&=\frac{\sum_{j=0}^{L}a_j\epsilon^j}
{1+\sum_{j=1}^{M}b_j\epsilon^j},
\qquad L+M=N,\\
\omega^2_{[L/M]}&=P_{L/M}(1).
\end{aligned}
\end{equation}
with $L\simeq M$. All WKB frequencies reported below use this direct Pad\'e prescription; Borel--Pad\'e resummation is not used.

For the present non-rational potentials, the barrier maximum is determined numerically and the required tortoise-coordinate derivatives are generated from local Taylor expansions of $V(r)$ and $A(r)$ around that maximum~\cite{Konoplya:2026wkb}. We calculate the WKB--Pad\'e sequence to high order and assess its internal reliability from both the diagonal or near-diagonal Pad\'e cluster and the Ces\`aro means of that sequence~\cite{Konoplya:2026rjh}. Stabilization of the Ces\`aro means is used as a convergence diagnostic; if it is not observed within the available orders, the corresponding mode is identified as numerically unresolved rather than assigned a precision estimate.

This method is applied to the scalar, electromagnetic, and Dirac potentials. As usual, it is most reliable for single-peak barriers and low-lying modes, particularly when $\ell\gtrsim n$; the low-multipole and overtone cases are therefore checked below by the independent time-domain and Chebyshev calculations.
}

\section{Time-domain integration and Prony analysis}

{The WKB--Pad\'e sequence supplies an internal convergence diagnostic but not an independent solution of the wave equation. We therefore evolve representative perturbations in the time domain, concentrating on the low-multipole and strongly deformed cases in which the WKB hierarchy is least robust. We use the characteristic integration scheme of Gundlach, Price, and Pullin~\cite{Gundlach:1994jp} and extract the ringdown frequencies by a Prony fit; see also the review~\cite{Konoplya:2011qq}. Related implementations have been widely used to determine dominant quasinormal modes and test perturbative stability~\cite{Konoplya:2005et,Dubinsky:2024mwd,Skvortsova:2025cah,Konoplya:2013sba,Lutfuoglu:2025bsf,Stuchlik:2025mjj,Skvortsova:2023zca,Konoplya:2014lha,Lutfuoglu:2025mqa,Dubinsky:2024gwo,Cuyubamba:2016cug}.} For each perturbation sector, one promotes the radial master function to a time-dependent field and solves
\begin{equation}\label{eq:time-domain-wave}
\frac{\partial^2 \Psi}{\partial t^2} - \frac{\partial^2 \Psi}{\partial r_*^2} + V(r)\Psi = 0,
\end{equation}
where $V(r)$ is one of the effective potentials $V_{\rm sc}$, $V_{\rm em}$, or $V_D^{\pm}$.

Introducing the null coordinates
\begin{equation}
u = t-r_*, \qquad v = t+r_*,
\end{equation}
Eq.~\eqref{eq:time-domain-wave} becomes
\begin{equation}\label{eq:time-domain-null}
4\,\frac{\partial^2 \Psi}{\partial u\partial v} + V\bigl(r(u,v)\bigr)\Psi = 0.
\end{equation}
On a uniform null grid with step size $\Delta$, let $S=(u,v)$, $E=(u,v+\Delta)$, $W=(u+\Delta,v)$, and $N=(u+\Delta,v+\Delta)$ denote the south, east, west, and north points of a null rectangle. The Gundlach et al. integration scheme is then
\begin{equation}\label{eq:gpp-update}
\Psi_N = \Psi_W + \Psi_E - \Psi_S - \frac{\Delta^2}{8}V_S\bigl(\Psi_W + \Psi_E\bigr) + \mathcal{O}(\Delta^4),
\end{equation}
where $V_S$ denotes the potential evaluated inside the null cell after reconstructing $r$ from the relation $r_*=(v-u)/2$. Since the tortoise coordinate for the beyond-Horndeski metric is defined only implicitly through $dr_*/dr=A(r)^{-1}$, in practice one tabulates the monotonic map $r\mapsto r_*$ numerically in the exterior region and obtains $r(u,v)$ by interpolation or root finding at each step.

The characteristic evolution requires initial data on two null segments. A convenient choice is a Gaussian pulse on one initial ray and trivial data on the other,
\begin{equation}\label{eq:time-domain-initial}\begin{array}{rcl}
\Psi(u_0,v) &=& \exp\left[-\dfrac{(v-v_c)^2}{2\sigma^2}\right],
\\
\Psi(u,v_0)&=&\Psi(u_0,v_0)=0,\end{array}
\end{equation}
with $v_c$ and $\sigma$ fixing the location and width of the initial perturbation. The precise shape of the localised initial pulse is unimportant for the late-time ringing: after the prompt response has passed, the waveform at a fixed observer position $r_*^{\rm obs}$ is governed by the quasinormal spectrum of the background.

At fixed $r_*^{\rm obs}$ the signal typically exhibits three stages: an initial burst determined by the chosen data, an intermediate quasinormal ringing regime, and a late-time tail. The quasinormal frequencies are extracted from the intermediate interval, where the waveform can be approximated as a sum of damped exponentials.
\begin{figure*}[tbp]
    \centering
    \begin{minipage}[t]{0.32\textwidth}
        \centering
        \includegraphics[width=\linewidth]{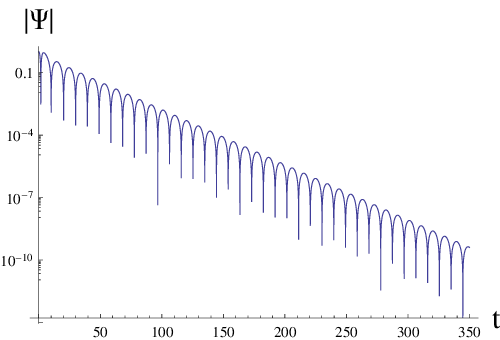}\\[-0.3em]
        {\small (a) Electromagnetic, $p_2=-0.2$}
    \end{minipage}\hfill
    \begin{minipage}[t]{0.32\textwidth}
        \centering
        \includegraphics[width=\linewidth]{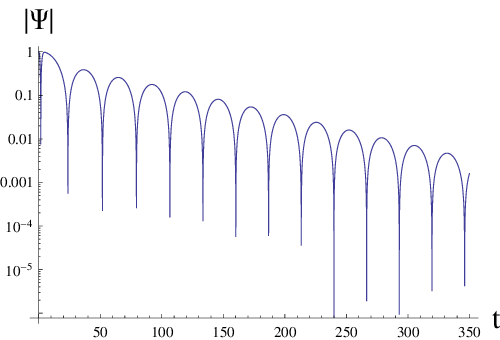}\\[-0.3em]
        {\small (b) Electromagnetic, $p_2=-0.6$}
    \end{minipage}\hfill
    \begin{minipage}[t]{0.32\textwidth}
        \centering
        \includegraphics[width=\linewidth]{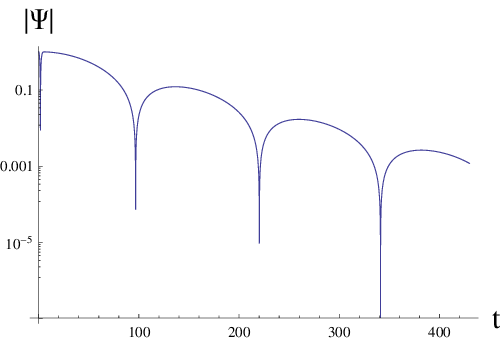}\\[-0.3em]
        {\small (c) Scalar, $p_1=0.06$, $p_2=-0.6$}
    \end{minipage}
    \caption{Representative time-domain profiles. Panels (a) and (b) show the semilogarithmic electromagnetic signal for $s=1$, $\ell=2$, and $p_1=0.01$. The WKB--Pad\'e frequencies $0.3294609381-0.0609704066i$ and $0.1183501198-0.0153824025i$ agree with the corresponding Prony estimates $0.329461-0.0609703i$ and $0.118333-0.0153724i$ at the levels of $10^{-7}$ and $10^{-5}$, respectively. Panel (c) shows the more delicate scalar monopole profile; contamination by the initial burst and late-time tail prevents a comparably precise Prony extraction, but the ringing is consistent with the independently determined frequency scale.}
    \label{fig:td-profiles}
\end{figure*}

Figure~\ref{fig:td-profiles}(a) illustrates this behavior for a representative electromagnetic signal with $s=1$, $\ell=2$, $p_1=0.01$, and $p_2=-0.2$. After the initial burst, the waveform enters a long ringing stage whose nearly linear envelope in the semilogarithmic plot is characteristic of exponential damping, while the repeated oscillations reflect the nonzero real part of the quasinormal frequency. The excellent agreement between the WKB--Pad\'e and Prony frequencies confirms that this interval is dominated by the fundamental electromagnetic mode and is therefore the appropriate window for the fit.

Figure~\ref{fig:td-profiles}(b) shows the analogous waveform for the same mode at $p_1=0.01$ and $p_2=-0.6$. In comparison with panel (a), the oscillations are slower and the damping is weaker, in agreement with the smaller values of ${\rm Re}(\omega)$ and $|{\rm Im}(\omega)|$. The continued close agreement between the WKB--Pad\'e and Prony frequencies shows that the time-domain extraction remains reliable also in this more weakly damped regime.

Figure~\ref{fig:td-profiles}(c) shows the corresponding $\ell=0$ scalar profile for the more delicate case $p_1=0.06$ and $p_2=-0.6$. The sixth-order WKB approximation gives $\omega=0.0248962-0.0143440i$, whereas the 500th- and 1000th-order values are $0.025519283782-0.015976980837i$ and $0.025598-0.015849i$, respectively. The independent Chebyshev/scattering calculation gives $\omega=0.0255615601-0.0157298838i$, within the slowly drifting high-order sequence. The time-domain waveform is compatible with this frequency scale, but contamination by the prompt response and tail prevents the Prony fit from providing comparable precision.

\begin{equation}\label{eq:prony-ansatz}
\Psi(t,r_*^{\rm obs}) \simeq \sum_{j=1}^{p} C_j e^{-i\omega_j (t-t_0)}.
\end{equation}
Here $p$ denotes the number of damped exponentials retained in the fit.
Sampling the signal at equally spaced times $t_n=t_0+n\Delta t$ gives a sequence
\begin{equation}\label{eq:prony-discrete}
x_n \equiv \Psi(t_n,r_*^{\rm obs}) = \sum_{j=1}^{p} C_j z_j^n,
\qquad z_j \equiv e^{-i\omega_j\Delta t}.
\end{equation}
The Prony method uses the fact that such a sequence satisfies a linear prediction relation,
\begin{equation}\label{eq:prony-linear}
x_{n+p} + \alpha_{p-1}x_{n+p-1} + \cdots + \alpha_1 x_{n+1} + \alpha_0 x_n = 0,
\end{equation}
with coefficients $\alpha_k$ determined from the sampled waveform, typically by solving an overdetermined linear system over the fitting window. One then forms the polynomial
\begin{equation}\label{eq:prony-poly}
z^p + \alpha_{p-1}z^{p-1} + \cdots + \alpha_1 z + \alpha_0 = 0,
\end{equation}
whose roots are the $z_j$. The quasinormal frequencies follow from
\begin{equation}\label{eq:prony-freq}
\omega_j = \frac{i}{\Delta t}\ln z_j,
\end{equation}
with the branch of the logarithm chosen so that ${\rm Re}(\omega_j)>0$, while we have ${\rm Im}(\omega_j)<0$ for the damped modes of interest. Once the roots are known, the amplitudes $C_j$ are recovered from the linear system~\eqref{eq:prony-discrete}.

In practical applications the main source of systematic uncertainty is the choice of fitting interval. The start time $t_0$ should be late enough that the prompt signal has decayed, but not so late that the waveform is already contaminated by power-law tails or numerical noise. It is therefore useful to vary the fitting window, the number of fitted exponentials $p$, and the observer location $r_*^{\rm obs}$, and to regard only the stable cluster of extracted frequencies as physically meaningful. In the present problem, such time-domain and Prony estimates provide a particularly valuable cross-check for the low-$\ell$ modes displayed below, where the WKB expansion is expected to be less reliable than for higher multipoles.

\section{Quasinormal modes}

{We now combine the methods above into a single spectral analysis. The WKB--Pad\'e calculation provides the full parameter dependence and overtone information, the Ces\`aro sequence identifies where those results are internally stable, and the time-domain profiles test selected modes directly. The independent Chebyshev calculation in Sec.~VII subsequently verifies the pole locations and supplies the systematic ADM-normalized comparison. Unless stated otherwise, the numerical calculations use $M=1$, so the raw values equal $M\omega$; the physical normalization $M_{\rm ADM}\omega$ is defined in Eq.~\eqref{eq:adm-frequency}.}

\subsection{Pad\'e sequence and reliability criterion}

\begin{figure*}[!t]
\centering
\resizebox{\linewidth}{!}{\includegraphics{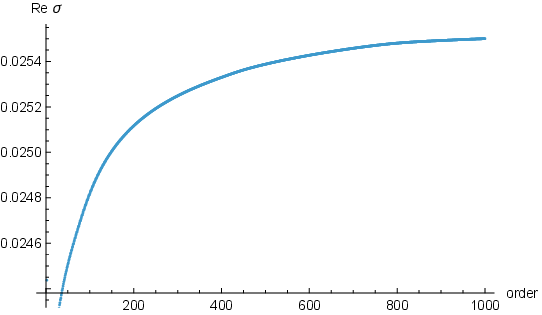}\includegraphics{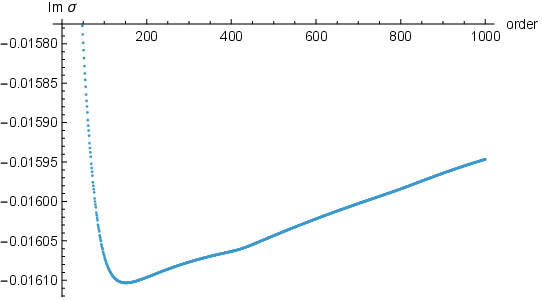}}
\caption{Ces\`aro means versus WKB order for $M=1$, $\ell=0$, $n=0$, $p_1=0.06$, $p_2=-0.6$. A monotonic regime is eventually reached, indicating slow convergence toward the more accurate value inferred from the time-domain profile in Fig.~\ref{fig:td-profiles}(c).}\label{fig:potCesaroC}
\end{figure*}

\begin{figure}[!t]
\centering
\resizebox{\linewidth}{!}{\includegraphics{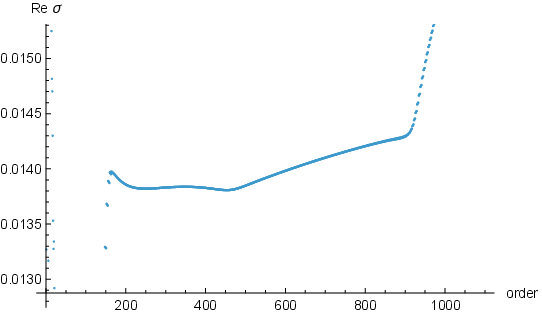}\includegraphics{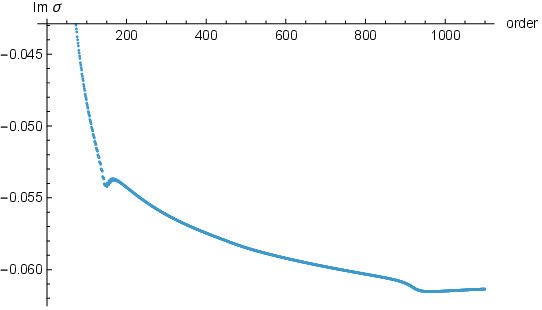}}
\caption{Ces\`aro means versus WKB order for $M=1$, $\ell=0$, $n=1$, $p_1=0.06$, $p_2=-0.6$. No clear stabilization is achieved within the displayed WKB orders.}\label{fig:potCesaroA}
\end{figure}
\begin{figure}[!t]
\centering
\resizebox{\linewidth}{!}{\includegraphics{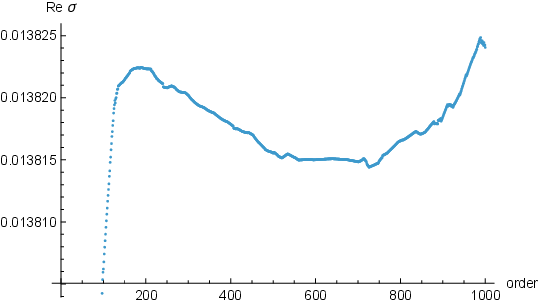}\includegraphics{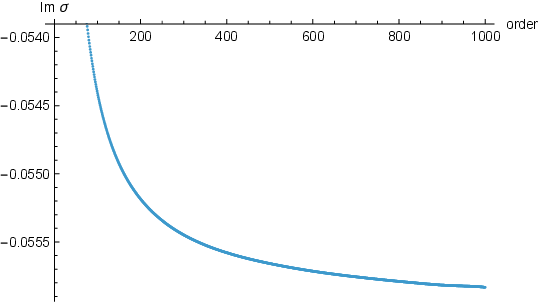}}
\caption{Ces\`aro means versus WKB order for $M=1$, $\ell=0$, $n=1$, $p_1=0.01$, $p_2=-0.6$. No clear stabilization is achieved within the displayed WKB orders.}\label{fig:potCesaroB}
\end{figure}
Two questions organize the numerical discussion: how the deformation parameters shift the spectrum, and whether a given WKB--Pad\'e sequence has reached a regime from which a reliable frequency can be inferred. {For the numerical analysis, we consider the sequence of quasinormal frequencies $\{\omega_N\}$ extracted from Pad\'e-resummed WKB expansions. Specifically, at each WKB order $N$, the frequency is defined via a diagonal or near-diagonal Pad\'e approximant,
\begin{equation}
\omega_N^2 = \mathcal{P}^{\tilde m}_{\tilde n}(1),
\qquad
\tilde m = \lfloor N/2 \rfloor, \quad \tilde n = N - \tilde m,
\label{eq:wkb-pade-sequence-new}
\end{equation}
where $\lfloor N/2 \rfloor$ denotes the integer part. Consequently, for even orders $N=2k$ one has $\tilde m=\tilde n=k$, while for odd orders $N=2k+1$ the approximant becomes slightly asymmetric, $\tilde m=k$, $\tilde n=k+1$.}

The behavior of the sequence $\{\omega_N\}$ reveals that straightforward convergence with increasing $N$ is generally absent. Instead, the Pad\'e-improved WKB results exhibit a characteristic oscillatory pattern, in which successive approximants fluctuate around a slowly varying central value. This type of non-monotonic convergence appears to be intrinsic to diagonal and near-diagonal Pad\'e constructions. In contrast, the direct WKB truncations $P_N(1)$ typically display poor convergence properties, with no comparable tendency toward stabilization even at high orders \cite{Matyjasek:2019eeu,HatsudaBorel}.

{To quantify the residual drift and suppress short-scale oscillations, it is useful to consider local averages of the sequence. Introducing the running mean
\begin{equation}
\overline{\omega}_{N,\Delta} = \frac{1}{\Delta+1} \sum_{s=0}^{\Delta} \omega_{N-s},
\label{eq:wkb-running-mean-new}
\end{equation}
one can smooth the rapid fluctuations by choosing $\Delta$ sufficiently large, while still retaining sensitivity to the slow evolution of the central value. A particularly important limiting case is obtained by averaging over the entire history of the sequence,
\begin{equation}
\sigma_N = \frac{1}{N+1} \sum_{k=0}^{N} \omega_k,
\label{eq:wkb-cesaro-new}
\end{equation}
which corresponds to the Ces\`aro mean. When $\sigma_N$ approaches a finite limit as $N \to \infty$, the original sequence is said to be Ces\`aro summable to that value \cite{HardyDivergent}. As emphasized in Ref.~\cite{Konoplya:2026rjh}, the Ces\`aro averages generally converge more slowly than the appropriate diagonal or near-diagonal Pad\'e approximants and are used here primarily as a conservative diagnostic: once the means stabilize, the associated high-order Pad\'e cluster supplies the more accurate finite-order frequency estimate.}

In the discussion below, we therefore interpret the numerical reliability of the WKB results through the behavior of the Ces\`aro means rather than through the spread of a few neighboring finite orders. When the Ces\`aro sequence enters a  monotonic regime, the corresponding mode will be regarded as numerically under control; when no such stabilization is seen within the available WKB orders, the quoted frequency should be treated with caution even if the raw Pad\'e values appear locally smooth. 

\subsection{Spectral response to the deformation parameters}

\begin{figure*}[!t]
    \centering
    \begin{minipage}[t]{0.49\textwidth}
        \centering
        \includegraphics[width=0.95\linewidth]{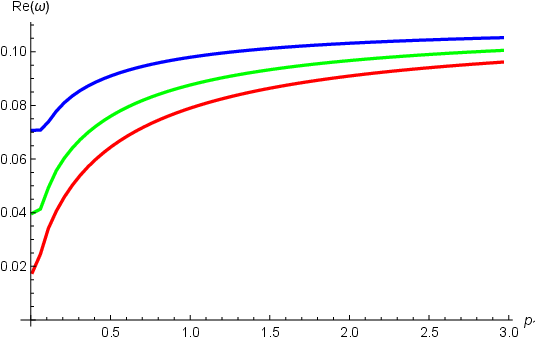}\\[-0.4em]
        {\small (a) ${\rm Re}(\omega)$}
    \end{minipage}\hfill
    \begin{minipage}[t]{0.49\textwidth}
        \centering
        \includegraphics[width=0.95\linewidth]{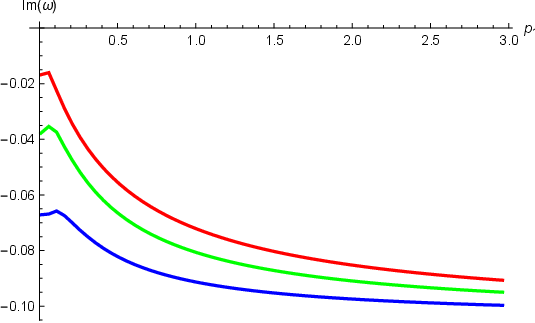}\\[-0.4em]
        {\small (b) ${\rm Im}(\omega)$}
    \end{minipage}
    \caption{Fundamental scalar quasinormal mode for $s=0$ and $\ell=0$ with $n=0$. Panel (a) shows the real part of $\omega$ and panel (b) the imaginary part, as functions of $p_1$ for $p_2=-0.2$ (blue), $p_2=-0.4$ (green), and $p_2=-0.6$ (red). The overall trend is toward larger oscillation frequency and stronger damping as $p_1$ grows, although the Chebyshev scan in Table~\ref{tab:cheb-scalar-scan} resolves a shallow low-$p_1$ structure for $p_2=-0.2$.}
    \label{fig:qnm-n0-panels}
\end{figure*}

\begin{figure*}[tbp]
    \centering
    \begin{minipage}[t]{0.49\textwidth}
        \centering
        \includegraphics[width=0.96\linewidth]{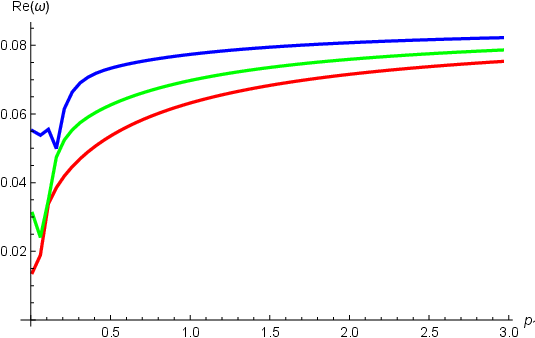}\\[-0.4em]
        {\small (a) ${\rm Re}(\omega)$}
    \end{minipage}\hfill
    \begin{minipage}[t]{0.49\textwidth}
        \centering
        \includegraphics[width=0.96\linewidth]{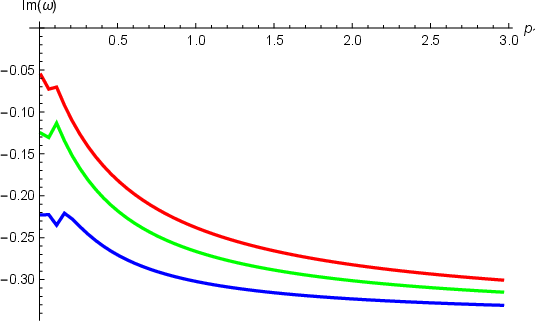}\\[-0.4em]
        {\small (b) ${\rm Im}(\omega)$}
    \end{minipage}
    \caption{First-overtone scalar quasinormal mode for $s=0$ and $\ell=0$ with $n=1$. Panel (a) shows the real part of $\omega$ and panel (b) the imaginary part, as functions of $p_1$ for $p_2=-0.2$ (blue), $p_2=-0.4$ (green), and $p_2=-0.6$ (red). The same qualitative parametric dependence as for the fundamental mode is preserved, but the overtone is more strongly damped.}
    \label{fig:qnm-n1-panels}
\end{figure*}

We begin with the scalar monopole, because it displays both the physical parameter dependence and the principal numerical limitation of the calculation. Figures~\ref{fig:qnm-n0-panels} and~\ref{fig:qnm-n1-panels} show, respectively, the fundamental mode and first overtone for $p_2=-0.2$ (blue), $-0.4$ (green), and $-0.6$ (red). Reading the two figures together separates a robust spectral trend from a convergence problem that is specific to the most strongly deformed, low-$p_1$ corner.

For the fundamental mode, the global evolution with $p_1$ is toward larger ${\rm Re}(\omega)$ and more negative ${\rm Im}(\omega)$ as the metric approaches its Schwarzschild limit. The higher-resolution Chebyshev calculation in Sec.~VII shows, however, that the scalar $p_2=-0.2$ branch contains a shallow nonmonotonic structure at very small $p_1$, below the resolution apparent in Fig.~\ref{fig:qnm-n0-panels}. At fixed $p_1$, decreasing $p_2$ lowers the oscillation frequency and weakens the damping.

The same figure also exposes the distinction between a smooth parameter dependence and a rapidly convergent numerical sequence. At $p_1=0.06$ and $p_2=-0.6$, the fundamental-mode Ces\`aro means in Fig.~\ref{fig:potCesaroC} drift only slowly, but eventually enter a monotonic regime directed toward the time-domain estimate in Fig.~\ref{fig:td-profiles}(c). The mode can therefore still be extracted, whereas a low-order WKB value cannot be assigned quantitative accuracy in this corner.

The first overtone preserves the same response to $p_1$ and $p_2$, but it lies at lower ${\rm Re}(\omega)$ and larger $|{\rm Im}(\omega)|$ than the fundamental mode. Its apparently regular curves must nevertheless be interpreted with greater care: for $p_2=-0.6$, the Ces\`aro means fail to stabilize within the available orders at both $p_1=0.01$ and $p_1=0.06$ [Figs.~\ref{fig:potCesaroB} and~\ref{fig:potCesaroA}]. Thus the qualitative deformation trend is clear, but individual frequencies in this narrow sector should not be quoted with precision unsupported by the high-order sequence.

\begin{figure*}[!t]
    \centering
    \begin{minipage}[t]{0.49\textwidth}
        \centering
        \includegraphics[width=0.96\linewidth]{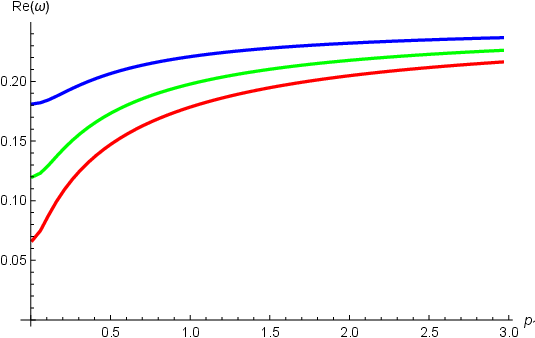}\\[-0.4em]
        {\small (a) ${\rm Re}(\omega)$}
    \end{minipage}\hfill
    \begin{minipage}[t]{0.49\textwidth}
        \centering
        \includegraphics[width=0.96\linewidth]{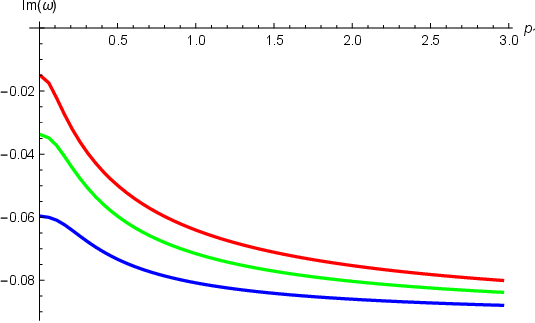}\\[-0.4em]
        {\small (b) ${\rm Im}(\omega)$}
    \end{minipage}
    \caption{Fundamental electromagnetic quasinormal mode for $s=1$ and $\ell=1$ with $n=0$. Panel (a) shows the real part of $\omega$ and panel (b) the imaginary part, as functions of $p_1$ for $p_2=-0.2$ (blue), $p_2=-0.4$ (green), and $p_2=-0.6$ (red). The real part increases with $p_1$, while the imaginary part stays negative and becomes more negative as $p_1$ grows.}
    \label{fig:qnm-em-panels}
\end{figure*}

\begin{figure*}[!t]
    \centering
    \begin{minipage}[t]{0.49\textwidth}
        \centering
        \includegraphics[width=0.96\linewidth]{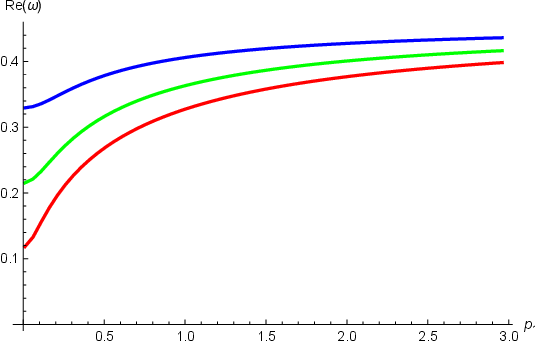}\\[-0.4em]
        {\small (a) ${\rm Re}(\omega)$}
    \end{minipage}\hfill
    \begin{minipage}[t]{0.49\textwidth}
        \centering
        \includegraphics[width=0.96\linewidth]{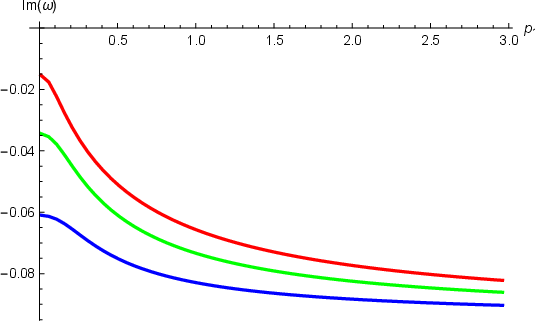}\\[-0.4em]
        {\small (b) ${\rm Im}(\omega)$}
    \end{minipage}
    \caption{Fundamental electromagnetic quasinormal mode for $s=1$ and $\ell=2$ with $n=0$. Panel (a) shows the real part of $\omega$ and panel (b) the imaginary part, as functions of $p_1$ for $p_2=-0.2$ (blue), $p_2=-0.4$ (green), and $p_2=-0.6$ (red). The real part increases with $p_1$, while the imaginary part stays negative and becomes more negative as $p_1$ grows.}
    \label{fig:qnm-em-l2-panels}
\end{figure*}

The electromagnetic sector tests whether this behavior is peculiar to the scalar monopole or is instead a property of the deformed geometry. We consider the fundamental modes with $\ell=1$ and $\ell=2$ in Figs.~\ref{fig:qnm-em-panels} and~\ref{fig:qnm-em-l2-panels}, using the same values and color convention for $p_2$.

Already at $\ell=1$, the electromagnetic spectrum reproduces the scalar trend: increasing $p_1$ moves the frequency upward and deeper into the damped half-plane, whereas decreasing $p_2$ lowers ${\rm Re}(\omega)$ and reduces $|{\rm Im}(\omega)|$. The $\ell=2$ data below retain the same ordering while shifting the oscillation frequency upward, as expected for the higher multipole. Unlike the scalar monopole overtone, neither electromagnetic sequence shows anomalously slow convergence over the displayed range, consistent with the greater reliability of WKB methods when the angular barrier is stronger.

The $\ell=2$ mode also supplies the cleanest independent validation of the resummed WKB calculation. At $p_1=0.01$ and $p_2=-0.2$, the WKB--Pad\'e frequency $\omega=0.3294609381-0.0609704066i$ agrees with the Prony value $0.329461-0.0609703i$ at the level of $10^{-7}$ [Fig.~\ref{fig:td-profiles}(a)]. For $p_2=-0.6$, the corresponding values, $0.1183501198-0.0153824025i$ and $0.118333-0.0153724i$, agree at the level of $10^{-5}$ [Fig.~\ref{fig:td-profiles}(b)]. The agreement persists even though the second signal is more weakly damped, supporting both the spectral trend and the quoted high-order frequency.

\begin{figure*}[!t]
    \centering
    \begin{minipage}[t]{0.49\textwidth}
        \centering
        \includegraphics[width=\linewidth]{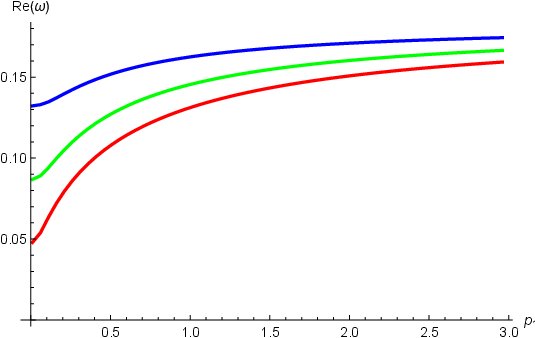}\\[-0.4em]
        {\small (a) ${\rm Re}(\omega)$}
    \end{minipage}\hfill
    \begin{minipage}[t]{0.49\textwidth}
        \centering
        \includegraphics[width=\linewidth]{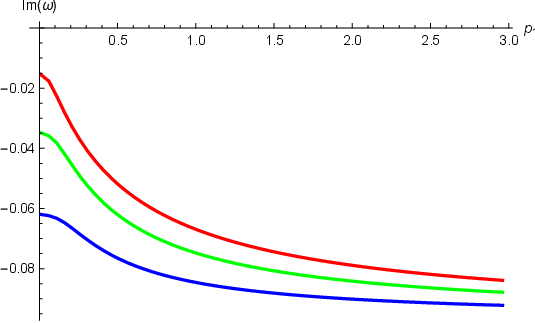}\\[-0.4em]
        {\small (b) ${\rm Im}(\omega)$}
    \end{minipage}
    \caption{Fundamental massless Dirac quasinormal mode for the lowest angular state $\ell=1/2$ (equivalently $\lambda=1$) with $n=0$. Panel (a) shows the real part of $\omega$ and panel (b) the imaginary part, as functions of $p_1$ for $p_2=-0.2$ (blue), $p_2=-0.4$ (green), and $p_2=-0.6$ (red). The real part increases with $p_1$, while the imaginary part remains negative and becomes more negative as $p_1$ grows.}
    \label{fig:qnm-dirac-panels}
\end{figure*}

The fundamental massless Dirac mode with $\ell=1/2$ ($\lambda=1$) provides a further check across field spin. Its spectrum is shown in Fig.~\ref{fig:qnm-dirac-panels} for the same three values of $p_2$.

The fermionic mode follows the same ordering as both bosonic sectors: ${\rm Re}(\omega)$ and $|{\rm Im}(\omega)|$ increase with $p_1$, while more negative $p_2$ produces slower oscillation and weaker damping. The high-order sequence remains regular throughout the plotted interval. The recurrence of this pattern for spins $0$, $1$, and $1/2$ indicates that the leading shift is controlled primarily by the background geometry rather than by details of a particular effective potential.

\subsection{Overtone hierarchy and comparison with sixth-order WKB}

\begin{figure*}[tbp]
    \centering
    \begin{minipage}[t]{0.48\textwidth}
        \centering
        \includegraphics[width=\linewidth]{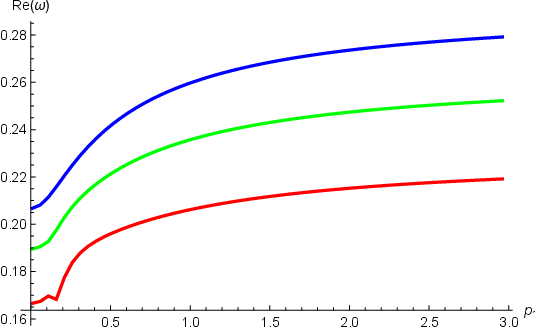}\\[-0.4em]
        {\small (a) ${\rm Re}(\omega)$}
    \end{minipage}\hfill
    \begin{minipage}[t]{0.48\textwidth}
        \centering
        \includegraphics[width=\linewidth]{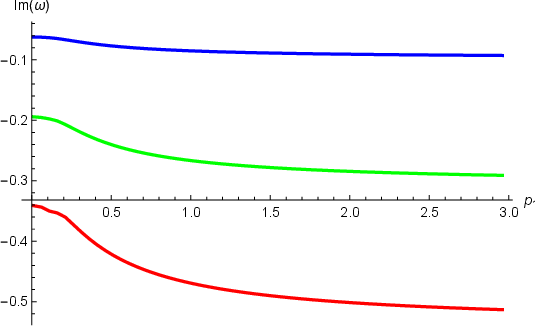}\\[-0.4em]
        {\small (b) ${\rm Im}(\omega)$}
    \end{minipage}
 \caption{\textbf{Comparison of scalar overtones at fixed $p_2=-0.2$ and $M=1$ for $s=0$ and $\ell=1$. Panel (a) shows the real part of $\omega$ and panel (b) the imaginary part for $n=0$ (blue), $n=1$ (green), and $n=2$ (red).}}
    
    \label{fig:qnm-overtones-panels}
\end{figure*}
\begin{figure*}[tbp]
\centering
\begin{minipage}[t]{0.49\textwidth}
\centering
\resizebox{\linewidth}{!}{\includegraphics{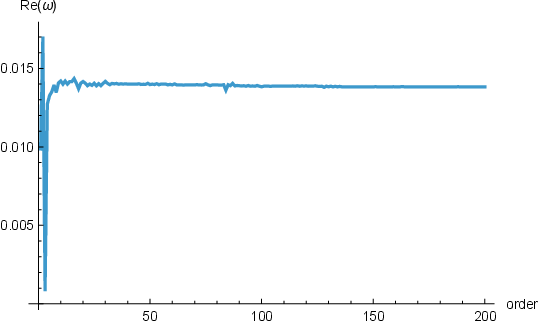}\includegraphics{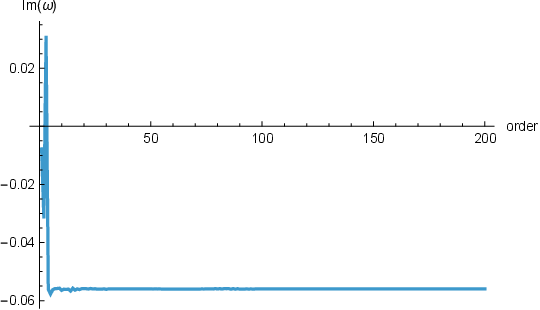}}\\[-0.3em]
{\small (a) $p_1=0.01$}
\end{minipage}\hfill
\begin{minipage}[t]{0.49\textwidth}
\centering
\resizebox{\linewidth}{!}{\includegraphics{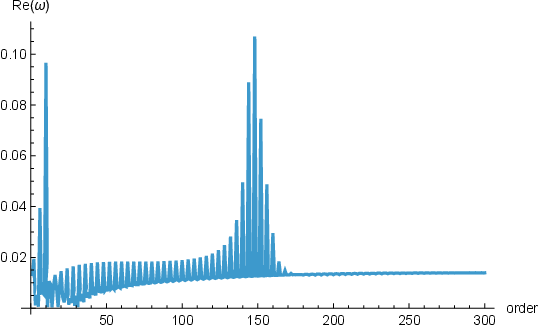}\includegraphics{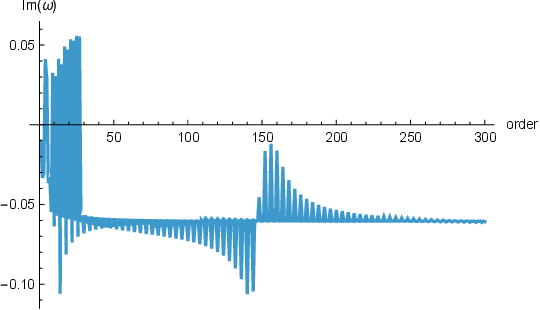}}\\[-0.3em]
{\small (b) $p_1=0.06$}
\end{minipage}
\caption{Order dependence of the Pad\'e-WKB scalar frequency for $p_2=-0.6$, $M=1$, $\ell=0$, and $n=1$. For $p_1=0.01$ the raw sequence is comparatively regular, although the corresponding Ces\`aro means have not fully stabilized and the sixth-order result differs from the high-order cluster by a few percent. For $p_1=0.06$ the order dependence is much stronger: the sixth-order value lies far from the high-order sequence, whose continuing drift prevents a precise error assignment.}
\label{fig:Convergence-panels}
\end{figure*}

To illustrate the overtone hierarchy, Fig. ~\ref{fig:qnm-overtones-panels} compares
the scalar quasinormal frequencies for $\ell=1$ and
$n=0,1,2$ at $p_2=-0.2$ and $M=1$. The convergence
properties of the more challenging scalar monopole
($\ell=0$) are examined separately in Fig. \ref{fig:Convergence-panels} and Table I.

\begin{table}[tbp]

\centering
\small
\resizebox{\columnwidth}{!}{%
\begin{tabular}{c c c}
\hline
WKB order & $p_1=0.01$  & $p_1=0.06$ \\
\hline
6   & $0.0135122 - 0.0562507\, i$ & $0.0393663 - 0.0365796\, i$ \\
7   & $0.0139419 - 0.0558639\, i$ & $0.0059011 - 0.0329113\, i$ \\
8   & $0.0134729 - 0.0558203\, i$ & $0.0059231 - 0.0547641\, i$ \\
9   & $0.0140747 - 0.0557294\, i$ & $0.0051789 + 0.0325976\, i$ \\
10  & $0.0141941 - 0.0564548\, i$ & $0.0965587 - 0.0637573\, i$ \\
50  & $0.0139704 - 0.0559757\, i$ & $0.0069307 - 0.0587957\, i$ \\
100 & $0.0138640 - 0.0559584\, i$ & $0.0106826 - 0.0596482\, i$ \\
150 & $0.0138271 - 0.0559693\, i$ & $0.0129657 - 0.0602800\, i$ \\
195 & $0.0138202 - 0.0559734\, i$ & $0.0135407 - 0.0604150\, i$ \\
196 & $0.0138195 - 0.0559647\, i$ & $0.0135405 - 0.0604150\, i$ \\
197 & $0.0138200 - 0.0559735\, i$ & $0.0132456 - 0.0543323\, i$ \\
198 & $0.0138198 - 0.0559640\, i$ & $0.0135817 - 0.0604223\, i$ \\
199 & $0.0138224 - 0.0559706\, i$ & $0.0136182 - 0.0601962\, i$ \\
200 & $0.0138196 - 0.0559644\, i$ & $0.0135815 - 0.0604222\, i$ \\
295 & $0.013818 - 0.055976\, i$ & $0.0138935 - 0.0604528\, i$ \\
296 & $0.013805 - 0.055967\, i$ & $0.0139531 - 0.0610855\, i$ \\
297 & $0.013818 - 0.055976\, i$ & $0.0138877 - 0.0604538\, i$ \\
298 & $0.013804 - 0.055972\, i$ & $0.0139612 - 0.0611505\, i$ \\
299 & $0.013819 - 0.055977\, i$ & $0.0138802 - 0.0604551\, i$ \\
300 & $0.013804 - 0.055971\, i$ & $0.0139690 - 0.0612158\, i$ \\
495 & $0.013816 - 0.056000\, i$ & $ 0.015861 - 0.062704\, i$ \\
496 & $0.013806 - 0.055947\, i$ & $0.014040 - 0.063977\, i$ \\
497 & $0.013810 - 0.056011\, i$ & $0.014092 - 0.060620\, i$ \\
498 & $0.013810 - 0.055940\, i$ & $0.014034 - 0.064023\, i$ \\
499 & $0.013799 - 0.056018\, i$ & $0.015921 - 0.062541\, i$ \\
500 & $0.013809 - 0.055944\, i$ & $0.014035 - 0.064018\, i$ \\
\hline
\end{tabular}%
}
\caption{Convergence in WKB orders: $p_2=-0.6$, $M=1$, $\ell=0$, $n=1$ scalar field perturbations.}
\label{tab:selected_mode}
\end{table}

The overtone plot exhibits the familiar hierarchy: at fixed $p_1$, increasing $n$ lowers ${\rm Re}(\omega)$ and increases $|{\rm Im}(\omega)|$, so the higher modes oscillate more slowly and decay more rapidly.


Figure~\ref{fig:Convergence-panels}, together with Table~\ref{tab:selected_mode}, makes the contrast within that sector explicit. For $p_1=0.01$, the raw Pad\'e-WKB values fluctuate around a comparatively narrow high-order cluster, although the Ces\`aro means still lack full stabilization. At $p_1=0.06$, the order dependence is far stronger: even the values between orders 200 and 500 continue to drift and fluctuate. In the latter case it is therefore meaningful to diagnose the sixth-order result as unreliable, but not to attach a precise percentage error to it before an independent limiting value has been established.

This distinction sharpens the comparison with Antoniou, Pappas, and Kanti~\cite{Antoniou:2025fpc}. Their sixth-order WKB calculation correctly captures the qualitative trend reproduced here: stronger deviations from Schwarzschild, obtained by decreasing $p_1$ or making $p_2$ more negative, lower the oscillation frequency and lengthen the damping time. The limitation is quantitative and mode dependent. For the scalar fundamental mode, the discrepancy reaches several percent and about nine percent in the delicate example tested against the time-domain signal in Fig.~\ref{fig:td-profiles}(c); for the first overtone, the sixth-order value can be separated from the high-order sequence by an order-unity amount while that sequence itself remains slowly convergent. Sixth-order WKB is therefore adequate for identifying the trend, and should improve toward the eikonal regime $\ell\gg n$, but it is not a controlled precision method for the low-multipole, small-$p_1$, negative-$p_2$ sector.

\section{Chebyshev pseudospectral calculation}

\begingroup

{The preceding WKB and time-domain calculations establish the spectral trends, but a frequency-domain boundary-value solution is needed to test the pole locations without either asymptotic expansion or waveform fitting. We therefore solve the scalar and electromagnetic master equations by Chebyshev pseudospectral collocation. This calculation is independent of the WKB--Pad\'e and Prony procedures; its converged eigenvalues subsequently serve only as initial seeds for the distinct scattering calculation of Sec.~VIII. Pseudospectral and matrix formulations of the QNM problem have been developed for a broad range of black-hole geometries~\cite{Jansen:2017oag,Lin:2016sch,Stashko:2026kja,Konoplya:2026cus,Konoplya:2025hgp,Konoplya:2024lch,Batic:2026mst,Fortuna:2020obg}, while compactified Chebyshev representations are particularly natural for the non-self-adjoint QNM operator of asymptotically flat spacetimes~\cite{Jaramillo:2020tuu}.}

We introduce the compact coordinate
\begin{equation}\label{eq:cheb-coordinate}
y=1-\frac{r_h}{r},
\qquad 0\leq y\leq1,
\end{equation}
so that $y=0$ is the event horizon and $y=1$ is spatial infinity. Directly approximating $\Psi$ is inefficient because the QNM boundary conditions are nonanalytic at both endpoints. We therefore factor out the known ingoing and outgoing behavior,
\begin{equation}\label{eq:cheb-factorization}
\Psi(r)=e^{i\omega r}
r^{i\omega\alpha}(r-r_h)^{-i\omega\beta}F(y),
\end{equation}
where
\begin{equation}
\beta=\frac{1}{A'(r_h)},
\qquad
\alpha=2M_{\rm ADM}+\beta.
\end{equation}
Near the horizon, Eq.~\eqref{eq:cheb-factorization} behaves as $e^{-i\omega r_*}$, whereas at infinity it behaves as $e^{+i\omega r_*}$ up to a constant phase. The residual function $F(y)$ is therefore regular at both endpoints.

For convenience, define
\begin{equation}
J\equiv\frac{dy}{dr}=\frac{(1-y)^2}{r_h},
\qquad
g(r)\equiv1+\frac{\alpha}{r}-\frac{\beta}{r-r_h},
\end{equation}
and write the bosonic potential as $V=A H$, with
\begin{equation}
H(r)=\frac{\ell(\ell+1)}{r^2}
+(1-s^2)\frac{A'(r)}{r}.
\end{equation}
After substituting Eq.~\eqref{eq:cheb-factorization} into the master equation and dividing by $AJ$, the equation for $F$ becomes
\begin{equation}\label{eq:cheb-quadratic-ode}
C_2F_{,yy}+(C_{10}+\omega C_{11})F_{,y}
+(C_{00}+\omega C_{01}+\omega^2C_{02})F=0,
\end{equation}
where
\begin{align}
C_2 &= AJ, \\
C_{10} &= A\frac{dJ}{dy}+A', \\
C_{11} &= 2iAg, \\
C_{00} &= -\frac{H}{J}, \\
C_{01} &= i\frac{Ag'+A'g}{J}, \\
C_{02} &= \frac{A^{-1}-Ag^2}{J}.
\end{align}
Although some terms in these expressions separately diverge at an endpoint, their combinations have finite limits. These limits are evaluated analytically and supply the first and last collocation rows. In particular, with $a_1=A'(r_h)$, $a_2=A''(r_h)/2$, $g_0=1+\alpha/r_h$, and $Q^2=-p_2/p_1^2$, the horizon row is
\begin{equation}\label{eq:cheb-horizon-row}
\begin{split}
0={}&(a_1-2i\omega)F_{,y}
+\Biggl[-\frac{\ell(\ell+1)}{r_h}-(1-s^2)a_1 \\
&+i\omega r_h\left(a_1g_0-\frac{a_2}{a_1}\right)
+2\omega^2r_h\left(g_0-\frac{a_2}{a_1^2}\right)\Biggr]F,
\end{split}
\end{equation}
while the infinity row is
\begin{equation}\label{eq:cheb-infinity-row}
\begin{split}
0={}&2i\omega F_{,y}-\frac{\ell(\ell+1)}{r_h}F\\
&+\frac{\omega^2}{r_h}
\left(8M_{\rm ADM}^2-2Q^2+2\beta r_h\right)F.
\end{split}
\end{equation}

We use the Chebyshev--Lobatto nodes
\begin{equation}
y_j=\frac{1-\cos(\pi j/N)}{2},
\qquad j=0,\ldots,N,
\end{equation}
and replace $F_{,y}$ and $F_{,yy}$ by the corresponding differentiation matrices. Equation~\eqref{eq:cheb-quadratic-ode} then becomes the quadratic matrix pencil
\begin{equation}\label{eq:cheb-pencil}
\left(M_0+\omega M_1+\omega^2M_2\right)\mathbf{F}=0.
\end{equation}
It is solved through the generalized linear eigenvalue problem
\begin{equation}\label{eq:cheb-linearization}
\begin{pmatrix}
0&I\\
-M_0&-M_1
\end{pmatrix}
\begin{pmatrix}\mathbf{F}\\ \omega\mathbf{F}\end{pmatrix}
=\omega
\begin{pmatrix}
I&0\\
0&M_2
\end{pmatrix}
\begin{pmatrix}\mathbf{F}\\ \omega\mathbf{F}\end{pmatrix}.
\end{equation}
Because the discretized operator is non-normal, a single-resolution spectrum contains spurious eigenvalues. We retain only modes that can be followed continuously in the parameters and that remain stable when $N$ is increased. At very high $N$, roundoff and differentiation-matrix conditioning eventually spoil the convergence; the quoted value is taken from the stable plateau rather than automatically from the largest matrix.

As a first check, the Schwarzschild scalar monopole and electromagnetic $\ell=2$ fundamental modes are found to be
\begin{align}
M\omega_{s=0,\ell=0} &= 0.1104549392-0.1048957155i,\\
M\omega_{s=1,\ell=2} &= 0.45759551163-0.09500442582i,
\end{align}
in agreement with the standard Schwarzschild spectrum~\cite{Leaver:1986gd,Berti:2009kk}.

\begin{table*}[t]

\centering
\small
\begin{tabular}{c c c}
\hline
$N$ or method & $p_2=-0.2$ & $p_2=-0.6$ \\
\hline
40  & $0.3430341101-0.0581468723i$ & $0.1183347569-0.0154840811i$ \\
60  & $0.3294514127-0.0612206260i$ & $0.1183494452-0.0153824972i$ \\
80  & $0.3294614971-0.0609672838i$ & $0.1183501226-0.0153824041i$ \\
100 & $0.3294608970-0.0609704119i$ & $0.1183501198-0.0153824025i$ \\
120 & $0.3294609375-0.0609704075i$ & $0.1183501198-0.0153824025i$ \\
WKB--Pad\'e & $0.3294609381-0.0609704066i$ & $0.1183501198-0.0153824025i$ \\
Prony & $0.329461-0.0609703i$ & $0.118333-0.0153724i$ \\
\hline
\end{tabular}
\caption{Convergence with Chebyshev order $N$ for the electromagnetic fundamental mode with $s=1$, $\ell=2$, $p_1=0.01$, and $M=1$. The last two rows reproduce the WKB--Pad\'e and time-domain/Prony values used in Figs.~\ref{fig:td-profiles}(a) and~\ref{fig:td-profiles}(b).}
\label{tab:cheb-convergence}
\end{table*}

Table~\ref{tab:cheb-convergence} directly checks the modes extracted from the two time-domain profiles. For $p_2=-0.2$, the Chebyshev sequence converges more slowly but reaches the WKB--Pad\'e value at the $10^{-9}$ level and the Prony value at the accuracy quoted from the waveform. For $p_2=-0.6$, agreement with WKB--Pad\'e is already at the $10^{-11}$ level by $N=100$; the residual difference from Prony remains of order $10^{-5}$, consistently with the time-domain fitting uncertainty discussed in Sec.~V.

\begin{table*}[t]

\centering
\small
\resizebox{\textwidth}{!}{%
\begin{tabular}{c cc cc cc}
\hline
$p_1$ & \multicolumn{2}{c}{$p_2=-0.2$} & \multicolumn{2}{c}{$p_2=-0.4$} & \multicolumn{2}{c}{$p_2=-0.6$} \\
& $M\omega$ & $M_{\rm ADM}\omega$ & $M\omega$ & $M_{\rm ADM}\omega$ & $M\omega$ & $M_{\rm ADM}\omega$ \\
\hline
0.03 & $0.070837-0.067146i$ & $0.441738-0.418721i$ & $0.040166-0.037949i$ & $0.460783-0.435345i$ & $0.016930-0.013507i$ & $0.282871-0.225678i$ \\
0.06 & $0.070800-0.066916i$ & $0.256153-0.242102i$ & $0.041184-0.035381i$ & $0.256822-0.220633i$ & $0.025562-0.015730i$ & $0.226322-0.139272i$ \\
0.10 & $0.073107-0.065747i$ & $0.187943-0.169021i$ & $0.048300-0.036858i$ & $0.200038-0.152649i$ & $0.032648-0.021135i$ & $0.186499-0.120732i$ \\
0.30 & $0.085185-0.074618i$ & $0.129787-0.113687i$ & $0.066663-0.054961i$ & $0.136473-0.112515i$ & $0.053237-0.042685i$ & $0.136860-0.109734i$ \\
1.00 & $0.097964-0.091306i$ & $0.113352-0.105649i$ & $0.087612-0.080637i$ & $0.115136-0.105970i$ & $0.078983-0.072110i$ & $0.116203-0.106091i$ \\
3.00 & $0.105328-0.099759i$ & $0.110843-0.104983i$ & $0.100626-0.095094i$ & $0.111164-0.105052i$ & $0.096302-0.090838i$ & $0.111429-0.105107i$ \\
10.0 & $0.108784-0.103282i$ & $0.110492-0.104904i$ & $0.107161-0.101716i$ & $0.110528-0.104912i$ & $0.105585-0.100197i$ & $0.110561-0.104919i$ \\
\hline
\end{tabular}%
}
\caption{Chebyshev pseudospectral frequencies for the fundamental scalar monopole, $s=0$, $\ell=0$, and $n=0$. The $M\omega$ columns retain the fixed-$M$ normalization used by the WKB and time-domain calculations, while $M_{\rm ADM}\omega$ gives the same eigenvalues in physical asymptotic-mass units.}
\label{tab:cheb-scalar-scan}
\end{table*}

\begin{table*}[t]

\centering
\small
\resizebox{\textwidth}{!}{%
\begin{tabular}{c cc cc cc}
\hline
$p_1$ & \multicolumn{2}{c}{$p_2=-0.2$} & \multicolumn{2}{c}{$p_2=-0.4$} & \multicolumn{2}{c}{$p_2=-0.6$} \\
& $M\omega$ & $M_{\rm ADM}\omega$ & $M\omega$ & $M_{\rm ADM}\omega$ & $M\omega$ & $M_{\rm ADM}\omega$ \\
\hline
0.01 & $0.329461-0.060970i$ & $5.504621-1.018691i$ & $0.215414-0.034406i$ & $6.982858-1.115314i$ & $0.118350-0.015382i$ & $5.695468-0.740261i$ \\
0.03 & $0.329914-0.061056i$ & $2.057339-0.380742i$ & $0.216809-0.034635i$ & $2.487229-0.397329i$ & $0.122101-0.015896i$ & $2.040056-0.265592i$ \\
0.06 & $0.331398-0.061346i$ & $1.198996-0.221949i$ & $0.221237-0.035412i$ & $1.379629-0.220829i$ & $0.132857-0.017610i$ & $1.176317-0.155918i$ \\
0.10 & $0.334662-0.062044i$ & $0.860349-0.159503i$ & $0.230288-0.037251i$ & $0.953758-0.154278i$ & $0.151045-0.021302i$ & $0.862826-0.121685i$ \\
0.30 & $0.358676-0.068923i$ & $0.546478-0.105011i$ & $0.281351-0.051108i$ & $0.575980-0.104628i$ & $0.224162-0.039683i$ & $0.576274-0.102018i$ \\
1.00 & $0.406007-0.082923i$ & $0.469782-0.095948i$ & $0.363165-0.073359i$ & $0.477256-0.096406i$ & $0.327418-0.065672i$ & $0.481710-0.096619i$ \\
3.00 & $0.436351-0.090386i$ & $0.459198-0.095118i$ & $0.416869-0.086185i$ & $0.460524-0.095210i$ & $0.398953-0.082349i$ & $0.461621-0.095284i$ \\
10.0 & $0.450671-0.093546i$ & $0.457750-0.095015i$ & $0.443948-0.092131i$ & $0.457895-0.095025i$ & $0.437419-0.090758i$ & $0.458032-0.095035i$ \\
\hline
\end{tabular}%
}
\caption{Chebyshev pseudospectral frequencies for the fundamental electromagnetic mode, $s=1$, $\ell=2$, and $n=0$, in both the fixed-$M$ and physical ADM-mass normalizations.}
\label{tab:cheb-em-scan}
\end{table*}

\begin{figure*}[t]
\centering
\includegraphics[width=0.90\textwidth]{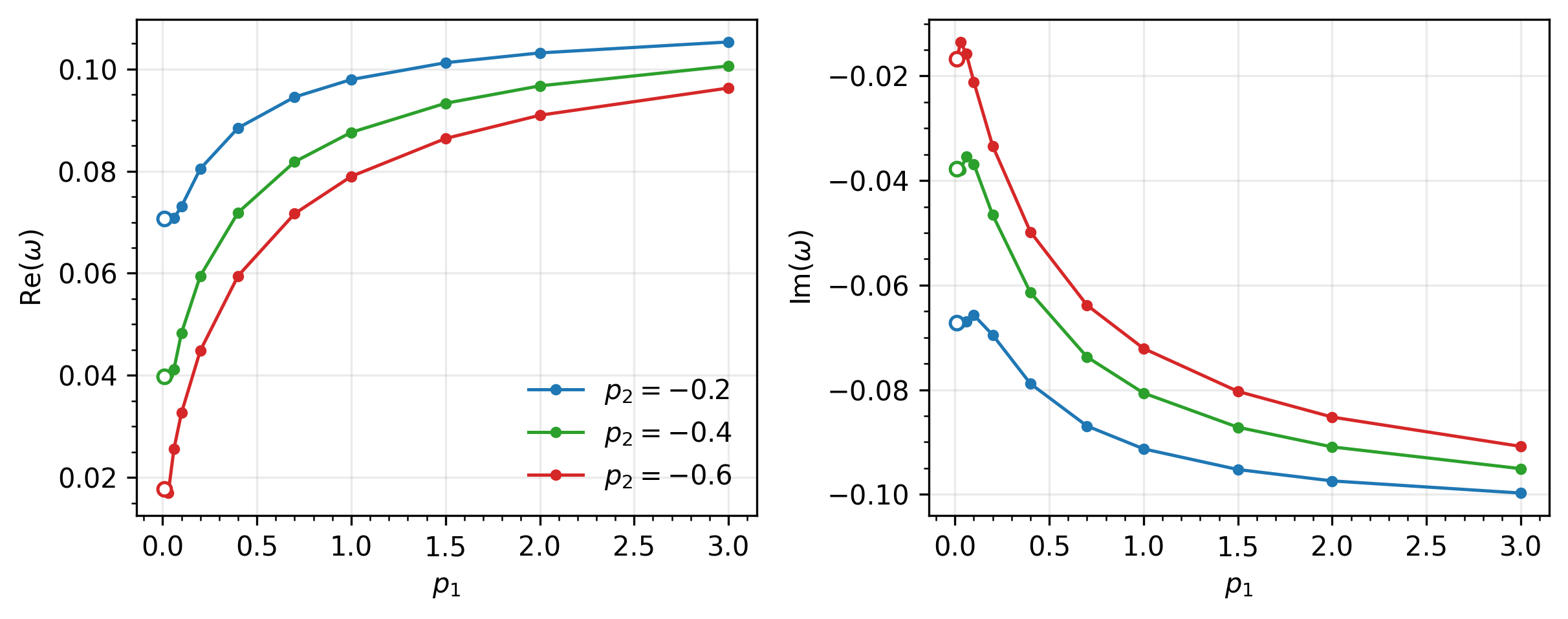}
\caption{Chebyshev pseudospectral frequencies of the fundamental scalar monopole. The points show the independently calculated grid in the fixed-$M$ normalization and the lines guide the eye. Blue, green, and red correspond to $p_2=-0.2$, $-0.4$, and $-0.6$, respectively. Open points mark parameter values whose double-precision resolution drift exceeds $10^{-7}$. {The corresponding ADM-normalized frequencies are given in Table~\ref{tab:cheb-scalar-scan}.}}
\label{fig:cheb-scalar}
\end{figure*}

\begin{figure*}[t]
\centering
\includegraphics[width=0.90\textwidth]{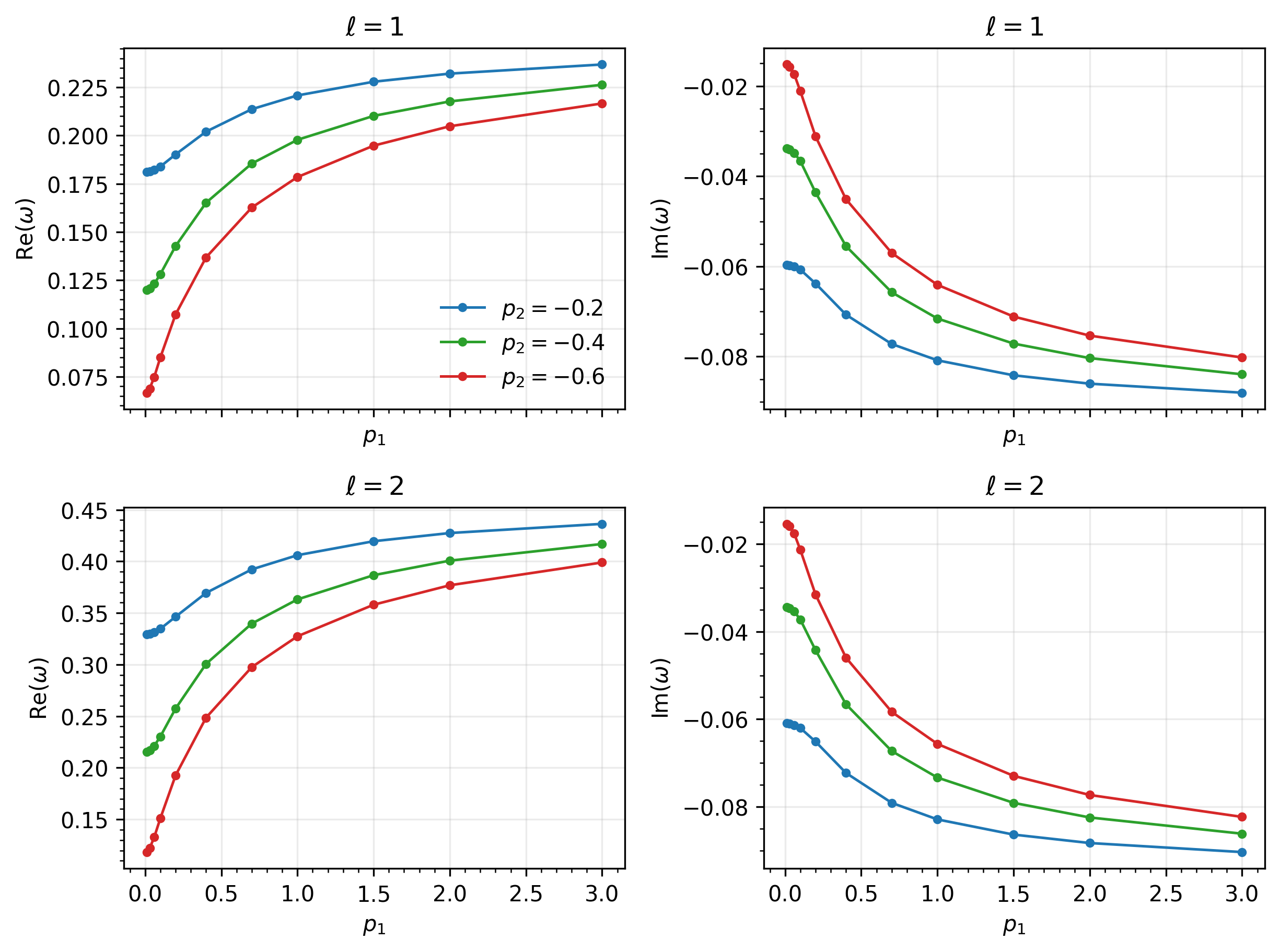}
\caption{Chebyshev pseudospectral frequencies of the fundamental electromagnetic modes with $\ell=1$ (upper panels) and $\ell=2$ (lower panels), using the same color convention as Fig.~\ref{fig:cheb-scalar}. {The plotted values use the fixed-$M$ normalization; Table~\ref{tab:cheb-em-scan} also lists the $\ell=2$ results as $M_{\rm ADM}\omega$.}}
\label{fig:cheb-em}
\end{figure*}

{The grid underlying Figs.~\ref{fig:cheb-scalar} and~\ref{fig:cheb-em} contains 99 candidate eigenvalues: the scalar monopole and the electromagnetic $\ell=1,2$ fundamental modes at eleven values of $p_1$ between $0.01$ and $3$, for each of $p_2=-0.2,-0.4,-0.6$. For every parameter pair we calculate the spectrum for $N=40,50,60,70,80$ and select the center of the three-resolution cluster with the smallest maximum drift. If that drift exceeds $5\times10^{-8}$, the sequence is extended through $N=120$. The selected order and the residual resolution drift are retained as numerical diagnostics. The $p_1=10$ entries in Tables~\ref{tab:cheb-scalar-scan} and~\ref{tab:cheb-em-scan} are separate large-$p_1$ checks obtained with the same method.}

{The two normalizations reveal different aspects of the parameter dependence. At fixed $M$, the electromagnetic scan confirms the smooth WKB--Pad\'e trend: increasing $p_1$ moves the mode toward Schwarzschild, whereas making $p_2$ more negative lowers both the oscillation frequency and the damping magnitude. The scalar scan has the same global behavior but retains the shallow nonmonotonic feature at very small $p_1$. In ADM units, however, the branches with $p_1\gtrsim1$ cluster closely around the Schwarzschild values: for example, at $p_1=3$ the scalar frequencies lie between $0.110843-0.104983i$ and $0.111429-0.105107i$, compared with the Schwarzschild value $0.110455-0.104896i$. The strongly deformed small-$p_1$ sector is qualitatively different because $M_{\rm ADM}/M$ becomes large. There $M_{\rm ADM}\omega$ can substantially exceed the Schwarzschild value and its dependence on $p_2$ is nonmonotonic; for the electromagnetic quadrupole at $p_1=0.01$, the three real parts are $5.50$, $6.98$, and $5.70$. Thus the apparent softening of the raw fixed-$M$ spectrum should not be interpreted as a universal softening in physical mass units. The scalar configurations with $p_1=0.01$ remain omitted from the tabulated high-precision scan because their double-precision collocation sequences do not develop sufficiently broad stable plateaus before differentiation-matrix conditioning dominates.}

For the delicate scalar configuration with $p_1=0.06$ and $p_2=-0.6$, the stable plateau gives
\begin{equation}
\omega=0.0255615600-0.0157298839i,
\end{equation}
in agreement with the independent scattering-root calculation and within the slowly drifting high-order WKB sequence. This confirms that the discrepancy of the sixth-order value is a low-order WKB error rather than a time-domain artifact. The scalar first overtone is substantially less robust: no resolution-independent Chebyshev cluster was found up to $N=160$ at $p_2=-0.6$ for either $p_1=0.01$ or $p_1=0.06$. We therefore do not quote pseudospectral overtone frequencies. This negative result is consistent with the sensitivity of overtones in non-self-adjoint black-hole spectral problems~\cite{Jaramillo:2020tuu} and reinforces the caution inferred from the Ces\`aro analysis.

The considerable enhancement of the quasinormal frequencies in ADM-mass units in the small-$p_1$ regime can be understood analytically. Indeed, for $p_1r\ll 1$, the metric function (21) takes the form
\begin{equation}
A(r)=1+p_2-\frac{2M}{r}
-\frac{p_2p_1^2r^2}{3}
+\mathcal{O}(p_1^4r^4).
\end{equation}
Introducing $c\equiv 1+p_2>0$, one finds that, in the limit $p_1\to 0$, the event horizon approaches $r_h=2M/c$, whereas the ADM mass grows without bound. Consequently,
\begin{equation}
\frac{r_h}{M_{\mathrm{ADM}}}
\simeq \frac{2M}{c}
\left(M-\frac{\pi p_2}{4p_1}\right)^{-1}
\longrightarrow 0.
\end{equation}
Thus, the characteristic length scale of the near-horizon geometry becomes much smaller than that associated with the total mass measured at spatial infinity. This separation of scales explains the considerable enhancement of the ADM-normalized quasinormal frequencies.

Moreover, for the scalar monopole ($\ell=0$), a simple analytical estimate of the quasinormal frequencies can be obtained in this regime. Introducing the rescaled radial and tortoise coordinates
\begin{equation}
R=cr,\qquad R_*=c^2r_*,
\end{equation}
one finds that the leading-order wave equation reduces to
\begin{equation}
\frac{d^2\Psi}{dR_*^2}
+\left[
\frac{\omega^2}{c^4}
-\left(1-\frac{2M}{R}\right)\frac{2M}{R^3}
\right]\Psi=0,
\end{equation}
which coincides with the Schwarzschild scalar monopole equation upon rescaling the frequency. This suggests the asymptotic relation
\begin{equation}
\omega\simeq (1+p_2)^2\omega_{\mathrm{Sch}}.
\end{equation}
For instance, at $p_2=-0.2$, this relation yields
$M\omega\simeq 0.070691-0.067133i$,
which is in excellent agreement with the numerical value
$M\omega=0.070837-0.067146i$ obtained at $p_1M=0.03$
(see Table III). Strictly speaking, the small-$p_1$ expansion is not uniform throughout the exterior region, since it breaks down at sufficiently large distances. Nevertheless, the above estimate provides a simple analytical interpretation of the numerical results in the strongly deformed regime.

\subsection{First electromagnetic overtones for the representative deformation}

{Reference~\cite{Konoplya:2022pbc} showed that a deformation confined mainly to the near-horizon region can leave the fundamental mode almost unchanged while producing a large separation of the first few overtones from their Schwarzschild values. To test whether the same mechanism is present here, we consider the electromagnetic quadrupole for the background represented by the blue curves in Fig.~\ref{fig:effective-potentials}: $M=1$, $p_1=0.3$, and $p_2=-0.2$, for which $M_{\rm ADM}=1.5235987756$ and $p_1M_{\rm ADM}=0.4570796$. This sector is preferable to the scalar monopole for a high-overtone test because its angular barrier gives a substantially better-conditioned pseudospectral sequence.}

{The factored Chebyshev problem was recomputed with multiprecision arithmetic, and each mode was followed continuously from the Schwarzschild limit to the selected deformation. Successive calculations through $N=75$ stabilize the first five rows strongly; the last two rows are quoted to fewer digits in accordance with their slower resolution convergence. The Schwarzschild column in Table~\ref{tab:cheb-em-overtones} is the Frobenius benchmark of Ref.~\cite{Konoplya:2022pbc}, converted from its convention $2M=1$. The final two columns give the percentage changes of ${\rm Re}(M_{\rm ADM}\omega_n)$ and $|{\rm Im}(M_{\rm ADM}\omega_n)|$ relative to this benchmark.}

\begin{table*}[t]
\centering
\small
\resizebox{0.85\textwidth}{!}{%
\begin{tabular}{c c c c c c}
\hline
$n$ & $M\omega_n^{\rm Sch}$ & $M\omega_n$ & $M_{\rm ADM}\omega_n$ & $\Delta_{\rm Re}$ & $\Delta_{|\rm Im|}$ \\
\hline
0 & $0.4575955-0.0950045i$ & $0.358675569-0.068922777i$ & $0.546477658-0.105010658i$ & $19.42\%$ & $10.53\%$ \\
1 & $0.4365425-0.2907100i$ & $0.345466562-0.210121891i$ & $0.526352431-0.320141456i$ & $20.57\%$ & $10.12\%$ \\
2 & $0.4011865-0.5015875i$ & $0.322967257-0.360127852i$ & $0.492072518-0.548690355i$ & $22.65\%$ & $9.39\%$ \\
3 & $0.3625950-0.7301985i$ & $0.29734324-0.52045274i$ & $0.45303179-0.79296115i$ & $24.94\%$ & $8.60\%$ \\
4 & $0.3287365-0.9716095i$ & $0.2730662-0.6884340i$ & $0.4160433-1.0488973i$ & $26.56\%$ & $7.96\%$ \\
5 & $0.3014930-1.2197150i$ & $0.251015-0.860375i$ & $0.382447-1.310866i$ & $26.85\%$ & $7.47\%$ \\
6 & $0.2798450-1.4707930i$ & $0.2295-1.0334i$ & $0.3497-1.5745i$ & $24.96\%$ & $7.05\%$ \\
\hline
\end{tabular}%
}
\caption{The fundamental mode and first six electromagnetic overtones for $s=1$ and $\ell=2$. The Schwarzschild frequencies are the Frobenius values of Ref.~\cite{Konoplya:2022pbc} in physical-mass units, while the deformed frequencies are shown in both the direct fixed-$M$ normalization and the physical ADM-mass normalization. The relative differences $\Delta_{\rm Re}$ and $\Delta_{|\rm Im|}$ compare $M_{\rm ADM}\omega_n$ with the corresponding Schwarzschild frequency. The reduced number of digits in the last two rows reflects their slower convergence with Chebyshev order.}
\label{tab:cheb-em-overtones}
\end{table*}

{Table~\ref{tab:cheb-em-overtones} does not show an overtone outburst for this parameter choice. In ADM-mass units the real part of the fundamental mode already differs from Schwarzschild by $19.4\%$, and the difference increases gradually to about $27\%$ rather than undergoing a sudden amplification. The relative change in the damping magnitude actually decreases from $10.5\%$ for the fundamental mode to about $7\%$ at $n=6$. This behavior is consistent with Fig.~\ref{fig:effective-potentials}: the blue electromagnetic barrier is visibly displaced and higher near its maximum, so the deformation is not restricted to a thin near-horizon region and the fundamental mode is not Schwarzschild-like. Thus the condition responsible for the outburst in Ref.~\cite{Konoplya:2022pbc} is not realized by the representative blue background; the present sequence instead displays a smooth, mode-dependent spectral shift.}

The stable Chebyshev frequencies are used as independent pole locations in the excitation-factor calculation of the next section. This separation is useful: the pseudospectral eigenvalue problem determines the QNM frequency and eigenfunction, while the scattering normalization required for $B_n$ is fixed by the Jost solutions and their Wronskian derivative.

\endgroup

\section{Quasinormal excitation factors}

\begingroup

{With the bosonic pole locations independently established, we turn from the position of each pole to its residue. The spectrum fixes the natural ringing scales, whereas the frequency-domain Green-function residue controls the weight with which a mode contributes to a response~\cite{Leaver:1986gd,Andersson:1995,Berti:2006wq}. Separating the excitation coefficient into an initial-data-independent excitation factor and a source-dependent integral, we calculate the former.}

Let $\Psi_{\rm in}(r,\omega)$ denote the solution normalized to be purely ingoing at the event horizon. At spatial infinity it has the form
\begin{equation}\label{eq:excitation-asymptotics}
\Psi_{\rm in}\sim
A_{\rm in}(\omega)e^{-i\omega r_*}
+A_{\rm out}(\omega)e^{+i\omega r_*}.
\end{equation}
A quasinormal frequency $\omega_n$ is a zero of $A_{\rm in}$. For the normalization in Eq.~\eqref{eq:excitation-asymptotics}, the corresponding excitation factor is~\cite{Berti:2006wq,Silva:2024}
\begin{equation}\label{eq:excitation-factor}
B_n=\left.
\frac{A_{\rm out}(\omega)}
{2\omega\,dA_{\rm in}/d\omega}
\right|_{\omega=\omega_n}.
\end{equation}
The excitation coefficient additionally contains an integral over the initial data or source. A large $|B_n|$ does not by itself imply a large observable waveform amplitude: the excitation factor depends on the normalization of the asymptotic radial waves.

The phase of $B_n$, and for a damped mode also its magnitude, depends on the additive constant chosen for the tortoise coordinate. To make comparisons unambiguous, we impose the asymptotic convention
\begin{equation}\label{eq:tortoise-normalization}
r_*=r+2M_{\rm ADM}\ln\!\left(\frac{r}{2M_{\rm ADM}}\right)
+\mathcal{O}(r^{-1}),
\qquad r\rightarrow\infty.
\end{equation}
In the Schwarzschild limit this convention reproduces
$r_*=r+2M\ln(r/2M-1)$.

Numerically, we construct a Frobenius series for $\Psi_{\rm in}$ at the event horizon and inverse-radius series for the ingoing and outgoing Jost solutions at infinity. Direct integration on the real radial axis is exponentially ill-conditioned at complex quasinormal frequencies. We therefore continue the asymptotic solutions along the anti-Stokes contour
\begin{equation}
r(\rho)=r_m+\rho\,e^{-i\arg\omega_n},
\qquad \rho\geq 0,
\end{equation}
on which the leading exponentials are oscillatory, and match the three solutions through their Wronskians at the real radius $r_m$. This complex-contour strategy is closely related to the phase-amplitude/Pr\"ufer approach developed for excitation factors beyond general relativity~\cite{Silva:2024}. We use the converged Chebyshev eigenvalues of Sec.~VII as independent initial seeds, refine each pole by solving $A_{\rm in}(\omega_n)=0$, and evaluate $dA_{\rm in}/d\omega$ with a symmetric five-point stencil. The quoted digits were checked by varying the horizon and infinity series from 12 to 16 terms, the matching point, the contour length, and the frequency step.

\begin{table*}[t]

\centering
\small
\resizebox{\textwidth}{!}{%
\begin{tabular}{c c l c c c c c c}
\hline
$s$ & $\ell$ & Case & $p_1$ & $p_2$ & $M\omega_0$ & $M_{\rm ADM}\omega_0$ & $B_0$ & $\mathcal{R}_B$ \\
\hline
0 & 0 & Schwarzschild (literature) & --- & $0$ & $0.11045494-0.10489572i$ & $0.11045494-0.10489572i$ & $0.21235-0.05927i$ & $1$ \\
0 & 0 & Schwarzschild (this work) & --- & $0$ & $0.11045494-0.10489572i$ & $0.11045494-0.10489572i$ & $0.21235-0.05927i$ & $1.000$ \\
0 & 0 & Schwarzschild-limit check & $100$ & $-0.6$ & $0.10993802-0.10440398i$ & $0.11045609-0.10489597i$ & $0.21235-0.05927i$ & $1.000$ \\
0 & 0 & Moderate deformation & $1$ & $-0.2$ & $0.09796422-0.09130628i$ & $0.11335240-0.10564864i$ & $0.21493-0.05778i$ & $1.009$ \\
0 & 0 & Moderate deformation & $1$ & $-0.6$ & $0.07898298-0.07211001i$ & $0.11620283-0.10609105i$ & $0.21731-0.05575i$ & $1.018$ \\
0 & 0 & Strong deformation & $0.06$ & $-0.2$ & $0.07079978-0.06691599i$ & $0.25615317-0.24210164i$ & $0.32491-0.02155i$ & $1.477$ \\
0 & 0 & Strong deformation & $0.06$ & $-0.6$ & $0.02556156-0.01572988i$ & $0.22632158-0.13927207i$ & $0.42523+0.16558i$ & $2.070$ \\
\hline
1 & 2 & Schwarzschild (literature) & --- & $0$ & $0.45759551-0.09500443i$ & $0.45759551-0.09500443i$ & $0.12119+0.01864i$ & $1$ \\
1 & 2 & Schwarzschild (this work) & --- & $0$ & $0.45759551-0.09500443i$ & $0.45759551-0.09500443i$ & $0.12119+0.01864i$ & $1.000$ \\
1 & 2 & Moderate deformation & $1$ & $-0.2$ & $0.40600668-0.08292279i$ & $0.46978206-0.09594827i$ & $0.12078+0.02531i$ & $1.006$ \\
1 & 2 & Moderate deformation & $1$ & $-0.6$ & $0.32741765-0.06567166i$ & $0.48170958-0.09661870i$ & $0.11965+0.03261i$ & $1.011$ \\
1 & 2 & Strong deformation & $0.01$ & $-0.2$ & $0.32946094-0.06097041i$ & $5.50462128-1.01869137i$ & $-0.01585+0.01257i$ & $0.165$ \\
1 & 2 & Strong deformation & $0.01$ & $-0.6$ & $0.11835012-0.01538240i$ & $5.69546813-0.74026092i$ & $-3.144-16.830i$ & $139.6$ \\
\hline
\end{tabular}%
}
\caption{Fundamental quasinormal frequencies and excitation factors, using the tortoise-coordinate convention~\eqref{eq:tortoise-normalization}. Both the fixed-$M$ and ADM-mass frequency normalizations are displayed. Here $\mathcal{R}_B\equiv |B_0|/|B_0^{\rm Sch}|$ measures the residue magnitude relative to the corresponding Schwarzschild mode. The literature rows summarize the Schwarzschild benchmarks of Refs.~\cite{Andersson:1995,Berti:2006wq,Lo:2025njp}; the following rows are our complex-contour results. Chebyshev frequencies from Sec.~VII seed the scattering-root refinement satisfying $A_{\rm in}=0$.}
\label{tab:excitation-factors}
\end{table*}

{The Schwarzschild rows in Table~\ref{tab:excitation-factors} provide a direct normalization check for both fields. For the scalar monopole we obtain $B_0=0.212349-0.059275i$, consistent with the published result $B_0\simeq0.212-0.059i$~\cite{Berti:2006wq}; for the electromagnetic quadrupole, $B_0=0.121190+0.018640i$ agrees at the displayed precision with the generalized Sasaki--Nakamura data of Ref.~\cite{Lo:2025njp}. T

he additional configuration with $p_1=100$ and $p_2=-0.6$ approaches the scalar Schwarzschild residue to better than $10^{-5}$ in relative magnitude, thereby checking the limiting behavior from within the deformed family.}

{The ratio $\mathcal{R}_B$ makes the deformation effect substantially clearer than the complex numbers alone. At $p_1=1$, the residue magnitudes differ from Schwarzschild by only about one to two percent in both sectors. For the strongly deformed scalar cases at $p_1=0.06$, the magnitude is enhanced by factors $1.48$ and $2.07$ for $p_2=-0.2$ and $-0.6$, respectively. The electromagnetic mode at $p_1=0.01$ is more sensitive: its residue is suppressed to $0.165$ of the Schwarzschild magnitude for $p_2=-0.2$, but is enhanced by a factor of about $140$ for $p_2=-0.6$, accompanied by a large phase rotation. These comparisons use the same tortoise-coordinate convention throughout; an observable waveform amplitude still requires the source-dependent integral and therefore cannot be inferred from $B_0$ alone.}

\subsection{Dependence on the tortoise-coordinate normalization}

The large variations in $|B_0|$ in Table~\ref{tab:excitation-factors} merit a closer examination. Although the quasinormal frequencies are unaffected by an additive shift of the tortoise coordinate, this is not generally true of the excitation factors. Consider the transformation $r_*\to r_*+C$, where $C$ is a real constant. Reexpressing the asymptotic solution~\eqref{eq:excitation-asymptotics} in terms of the shifted coordinate gives
\begin{equation}
A_{\rm in}\to e^{i\omega C}A_{\rm in},\qquad
A_{\rm out}\to e^{-i\omega C}A_{\rm out}.
\end{equation}
Since $A_{\rm in}(\omega_n)=0$, the derivative of the incoming amplitude at the pole acquires the same factor as $A_{\rm in}$. It follows from Eq.~\eqref{eq:excitation-factor} that
\begin{equation}\label{eq:excitation-shift}
B_n\to e^{-2i\omega_n C} B_n.
\end{equation}
An overall change in the horizon normalization of $\Psi_{\rm in}$ cancels in this relation. Writing $\omega_n=\omega_{R,n}-i\gamma_n$, with $\gamma_n>0$, we obtain $|B_n|\to e^{-2\gamma_n C}|B_n|$. Thus, an additive shift, which affects only the phase of scattering amplitudes at real frequencies, can change the magnitude of the excitation factor exponentially at a quasinormal frequency. The corresponding change in the normalized radial mode functions compensates this factor in the full Green-function residue at fixed physical source and observer positions: for example, the asymptotic pole contribution contains the combination $B_n e^{i\omega_n(r_*+r_*')}$, which is invariant under the shift. Hence there is no corresponding change in the physical waveform for the same initial data.

To separate the large displacement of the potential barrier from changes in its shape, it is useful to introduce a second convention. Let $r_{\rm p}$ be the areal radius of the maximum of the effective potential for the given perturbation sector. We define
\begin{align}
x_{\rm p}&\equiv r_*(r_{\rm p}),\nonumber\\
\widetilde r_*&=r_*-x_{\rm p},\qquad
\widetilde r_*(r_{\rm p})=0,\label{eq:peak-normalization}\\
\widetilde B_n&=B_n e^{2i\omega_n x_{\rm p}}.\label{eq:peak-excitation}
\end{align}
This is a useful common reference for comparing single-barrier potentials, not a uniquely preferred physical normalization. In particular, the asymptotic prescription~\eqref{eq:tortoise-normalization} remains appropriate for specifying scattering amplitudes at infinity. The peak-normalized quantity $\widetilde B_n$ is independent of the original additive constant, once $r_{\rm p}$ has been specified. The offset $x_{\rm p}$ is evaluated directly from
\begin{align}\label{eq:tortoise-offset}
x_{\rm p}={}&r_{\rm p}+2M_{\rm ADM}\ln\!\left(\frac{r_{\rm p}}{2M_{\rm ADM}}\right)\nonumber\\
&-\int_{r_{\rm p}}^{\infty}\!\left[\frac{1}{A(r)}-1-\frac{2M_{\rm ADM}}{r}\right]dr,
\end{align}
where the integrand falls off as $r^{-2}$. The comparison in Fig.~\ref{fig:peak-normalization} illustrates the two conventions for the electromagnetic quadrupole at fixed $M$. Although in the asymptotic convention the peak can be situated at a large negative tortoise coordinate, this does not mean that it lies unusually close to the horizon. For example, at $p_1M=0.01$ and $p_2=-0.6$, one finds $r_h/M\simeq4.994$ and $r_{\rm p}/M\simeq7.500$, so that $r_{\rm p}/r_h\simeq1.502$, almost identical to the Schwarzschild value $3/2$. The large negative offset $x_{\rm p}/M\simeq-158.006$ instead reflects the matching of the near-black-hole geometry to the asymptotically flat region over a large radial scale. This scale separation is also responsible for the enhancement of the ADM-normalized frequencies discussed in Sec.~VII.

\begin{figure*}[tbp]
\centering
\includegraphics[width=0.98\textwidth]{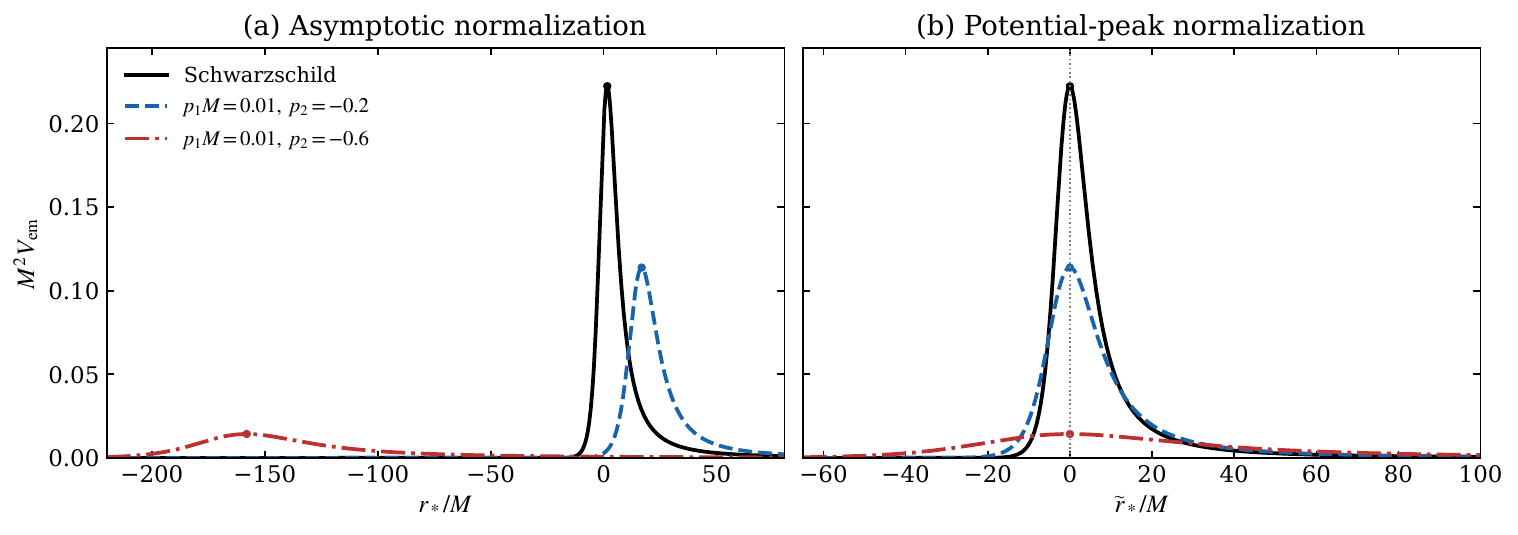}
\caption{Electromagnetic effective potential ($s=1$, $\ell=2$) for the Schwarzschild black hole (black solid curve) and the two strongly deformed configurations of Table~\ref{tab:excitation-factors} with $p_1M=0.01$ and $p_2=-0.2$ (blue dashed curve) or $p_2=-0.6$ (red dash-dotted curve), in units of the metric scale $M$. Panel (a) uses the asymptotic tortoise-coordinate normalization~\eqref{eq:tortoise-normalization}; the potential maxima are located at $r_*/M\simeq1.614$, $16.839$, and $-158.006$, respectively. Panel (b) uses the shifted coordinate~\eqref{eq:peak-normalization}, aligning all three maxima at the origin. Dots mark the potential maxima. The horizontal ranges differ between panels to display both the large coordinate displacement and the barrier shapes.}
\label{fig:peak-normalization}
\end{figure*}

\begin{table*}[tbp]
\centering
\small
\begin{tabular}{ccccrrrr}
\hline
$s$ & $\ell$ & $p_1M$ & $p_2$ & $r_{\rm p}/M$ & $x_{\rm p}/M$ & $\widetilde B_0$ & $\widetilde{\mathcal R}_B$ \\
\hline
0 & 0 & --- & $0$ & 2.667 & 0.469 & $0.23984-0.04080i$ & 1.000 \\
0 & 0 & 100 & $-0.6$ & 2.679 & 0.472 & $0.23984-0.04080i$ & 1.000 \\
0 & 0 & 1 & $-0.2$ & 2.992 & 0.427 & $0.23676-0.04283i$ & 0.989 \\
0 & 0 & 1 & $-0.6$ & 3.694 & 0.400 & $0.23348-0.04441i$ & 0.977 \\
0 & 0 & 0.06 & $-0.2$ & 3.336 & $-1.380$ & $0.26148-0.07003i$ & 1.113 \\
0 & 0 & 0.06 & $-0.6$ & 6.787 & $-28.246$ & $0.08965-0.16485i$ & 0.771 \\
\hline
1 & 2 & --- & $0$ & 3.000 & 1.614 & $-0.00977+0.16633i$ & 1.000 \\
1 & 2 & 1 & $-0.2$ & 3.357 & 1.697 & $-0.00226+0.16350i$ & 0.981 \\
1 & 2 & 1 & $-0.6$ & 4.131 & 1.941 & $0.00540+0.15993i$ & 0.960 \\
1 & 2 & 0.01 & $-0.2$ & 3.750 & 16.839 & $0.0851+0.1327i$ & 0.946 \\
1 & 2 & 0.01 & $-0.6$ & 7.500 & $-158.006$ & $0.0151-0.1317i$ & 0.796 \\
\hline
\end{tabular}
\caption{Companion to Table~\ref{tab:excitation-factors}: location of the potential maximum, its tortoise-coordinate offset in the asymptotic convention, and excitation factors transformed to the potential-peak convention~\eqref{eq:peak-excitation}. The ratio $\widetilde{\mathcal R}_B\equiv|\widetilde B_0|/|\widetilde B_0^{\rm Sch}|$ uses the Schwarzschild excitation factor in the same peak convention for each spin and multipole.}
\label{tab:peak-excitation}
\end{table*}

Table~\ref{tab:peak-excitation} gives the excitation factors transformed according to Eq.~\eqref{eq:peak-excitation}. The most strongly deformed electromagnetic configuration, for which the asymptotically normalized magnitude is nearly $140$ times its Schwarzschild counterpart, has $\gamma_0 M\simeq0.0153824$. The offset in this case contributes the factor $\exp(-2\gamma_0 x_{\rm p})\simeq129.2$ to $|B_0|$ relative to $|\widetilde B_0|$. In the peak-normalized convention, the excitation-factor magnitude is instead about $0.80$ of the corresponding Schwarzschild value; likewise, the suppression to $0.165$ at $p_2=-0.2$ becomes a ratio of about $0.95$. For the strongly deformed scalar monopole, the enhancements by factors $1.48$ and $2.07$ become ratios of $1.11$ and $0.77$, respectively. The pronounced changes in the asymptotically normalized excitation factors are therefore predominantly associated with the tortoise-coordinate offset, rather than with comparably large changes in the excitation factors referred to the potential maximum. This distinction is especially important for geometries with widely separated near-horizon and asymptotic mass scales. Neither convention by itself specifies the observable excitation amplitude, which also depends on the perturbation's initial data or source.


\begin{table*}[t]
\centering
\small
\begin{tabular}{cccccc}
\hline
$p_1M$ & $p_2$ & $M\omega_0$ & $M_{\rm ADM}\omega_0$
& $B_0$ & $\mathcal{R}_B$ \\
\hline
--- & $0$ &
$0.18296287-0.09698239i$ &
$0.18296287-0.09698239i$ &
$0.17955-0.03988i$ & 1.000 \\

100 & $-0.6$ &
$0.18210668-0.09652782i$ &
$0.18296484-0.09698270i$ &
$0.17955-0.03987i$ & 1.000 \\

1 & $-0.2$ &
$0.16250759-0.08458671i$ &
$0.18803422-0.09787356i$ &
$0.18177-0.03661i$ & 1.008 \\

1 & $-0.4$ &
$0.14545393-0.07479224i$ &
$0.19114963-0.09828891i$ &
$0.18308-0.03429i$ & 1.013 \\

1 & $-0.6$ &
$0.13118638-0.06693012i$ &
$0.19300651-0.09847020i$ &
$0.18381-0.03277i$ & 1.015 \\

0.01 & $-0.2$ &
$0.13226000-0.06197770i$ &
$2.20979528-1.03552119i$ &
$-0.01019+0.02799i$ & 0.162 \\

0.01 & $-0.4$ &
$0.08682944-0.03483037i$ &
$2.81465682-1.12905867i$ &
$2.270-1.256i$ & 14.11 \\

0.01 & $-0.6$ &
$0.04789955-0.01550977i$ &
$2.30511249-0.74639026i$ &
$-0.7574+27.5821i$ & 150.02 \\
\hline
\end{tabular}
\caption{Fundamental quasinormal frequencies and excitation factors
of the massless Dirac field ($\ell=1/2$, $\lambda=1$),
calculated for the effective potential $V_D^+$ using the
asymptotic tortoise-coordinate normalization
\eqref{eq:tortoise-normalization}.
The first row corresponds to the Schwarzschild black hole.}
\label{tab:dirac-excitation}
\end{table*}

\begin{table*}[t]
\centering
\small
\begin{tabular}{cccccc}
\hline
$p_1M$ & $p_2$ & $r_{\rm p}/M$ & $x_{\rm p}/M$
& $\widetilde B_0$ & $\widetilde{\mathcal{R}}_B$ \\
\hline
--- & $0$    & 2.420 & $-0.699$ &
$0.14290-0.07333i$ & 1.000 \\

100 & $-0.6$ & 2.432 & $-0.702$ &
$0.14290-0.07334i$ & 1.000 \\

1 & $-0.2$   & 2.710 & $-0.921$ &
$0.13938-0.07581i$ & 0.988 \\

1 & $-0.4$   & 3.017 & $-1.128$ &
$0.13706-0.07727i$ & 0.980 \\

1 & $-0.6$   & 3.337 & $-1.319$ &
$0.13559-0.07810i$ & 0.974 \\

0.01 & $-0.2$ & 3.061 & $13.478$ &
$0.11056-0.11334i$ & 0.986 \\

0.01 & $-0.4$ & 4.148 & $-40.460$ &
$0.04905-0.14691i$ & 0.964 \\

0.01 & $-0.6$ & 6.366 & $-168.302$ &
$-0.05638-0.13804i$ & 0.928 \\
\hline
\end{tabular}
\caption{Excitation factors of the fundamental Dirac mode
($\ell=1/2$) in the potential-peak normalization
\eqref{eq:peak-excitation}. Here $r_{\rm p}$ is the location
of the maximum of $V_D^+$, while $x_{\rm p}$ is its
tortoise-coordinate value in the asymptotic convention.}
\label{tab:dirac-peak-excitation}
\end{table*}

\subsection{Excitation factors of the Dirac field}

The calculation of excitation factors for the massless Dirac field
requires a somewhat different treatment of the near-horizon expansion
owing to the square-root dependence of the effective potential on
$A(r)$. Instead of constructing a Frobenius expansion directly for
the second-order master equation, we employ the equivalent system
of first-order Dirac equations. The purely ingoing solution near
the event horizon can be written as
\begin{align}
F &= e^{-i\omega r_*}\sqrt{r-r_h}
     \sum_{j=0}^{\infty} f_j(r-r_h)^j,\\
G &= e^{-i\omega r_*}
     \sum_{j=0}^{\infty} g_j(r-r_h)^j,
\end{align}
where the coefficients are determined recursively from the coupled
radial equations, with $g_0=1$.

At spatial infinity, the corresponding asymptotic amplitudes are
defined by
\begin{align}
F &\sim A_{\rm out}(\omega)e^{i\omega r_*},\\
G &\sim A_{\rm in}(\omega)e^{-i\omega r_*},
\end{align}
where subleading inverse-radius corrections to both wave components
are included in the numerical calculation. Thus, for the master
variable $Z_+=F+G$, the excitation factor is given directly by
Eq.~\eqref{eq:excitation-factor}. We use the same complex-contour
integration and asymptotic normalization of the tortoise coordinate
as for the bosonic fields. Representative results were additionally
verified by direct integration of the second-order master equation
with the potential $V_D^+$.

The results are presented in Table~\ref{tab:dirac-excitation}.
In the Schwarzschild limit, we recover the fundamental quasinormal
frequency reported in Ref.~\cite{CavalcanteDirac2021} and obtain
the excitation factor $B_0=0.179549-0.039875i$.
For moderate deviations from Schwarzschild, the magnitude of the
excitation factor changes by only a few percent. However, similarly
to the electromagnetic case, the strongly deformed configurations
exhibit much larger variations. In particular, for $p_1M=0.01$
and $p_2=-0.6$, the magnitude of the excitation factor exceeds
its Schwarzschild value by a factor of approximately $150$.

Reexpressing the excitation factors in the potential-peak
normalization, we find that these large deviations are almost
entirely removed, as shown in
Table~\ref{tab:dirac-peak-excitation}. In fact, the magnitudes
of the peak-normalized excitation factors differ from their
Schwarzschild value by less than eight percent throughout
the considered set of parameters. Thus, the Dirac sector
provides further evidence that the large enhancement of the
asymptotically normalized excitation factors is predominantly
associated with the displacement of the potential barrier
in the tortoise coordinate, rather than with a comparable
enhancement of the observable ringing amplitude.


\section{Conclusions}

{We have analyzed scalar, electromagnetic, and massless Dirac perturbations of the asymptotically flat parity-symmetric beyond-Horndeski black hole. After identifying the admissible exterior geometry and deriving the corresponding one-dimensional master equations, we calculated the spectrum with very-high-order WKB--Pad\'e expansions. Ces\`aro means of the resummed sequence were used as an internal reliability diagnostic, while characteristic time-domain evolution and Chebyshev collocation supplied independent checks of representative  modes.}

{The spectral trends are simplest in the fixed-$M$ normalization used for the direct calculations: increasing $p_1$ generally raises both the oscillation frequency and damping magnitude, whereas making $p_2$ more negative lowers them. The scalar monopole contains a shallow exception to strict monotonicity at very small $p_1$, while the fundamental Dirac mode follows the same overall trend as the bosonic sectors. Because the ADM mass varies along these scans, however, the raw frequencies do not by themselves compare black holes of the same physical mass. In terms of $M_{\rm ADM}\omega$, the mildly deformed solutions with $p_1\gtrsim1$ cluster near Schwarzschild, whereas the strongly deformed small-$p_1$ sector is amplified and becomes nonmonotonic. The overtone ordering is unaffected by this rescaling: higher overtones oscillate more slowly and decay more rapidly.}

{The numerical comparison also clarifies where each approximation is reliable. Over most of the parameter space, the high-order Pad\'e cluster, its Ces\`aro behavior, the time-domain signal, and the Chebyshev eigenvalue are mutually consistent. The exception is a narrow scalar-monopole sector at small $p_1$ and negative $p_2$, where convergence becomes slow and the first overtone does not develop a resolution-independent Chebyshev plateau within the available orders. The electromagnetic time-domain benchmarks agree with the WKB--Pad\'e values to about $10^{-7}$ and $10^{-5}$ in the two examples considered. In the delicate scalar case, the earlier sixth-order WKB approximation differs by about nine percent for the fundamental mode; for the first overtone it can be separated from the high-order sequence by an order-unity amount while that sequence is itself still drifting. Sixth-order WKB therefore captures the qualitative direction of the deformation but is not a controlled precision approximation in this low-multipole sector.}

{The multiprecision electromagnetic calculation further shows that the blue background of Fig.~\ref{fig:effective-potentials} does not exhibit the near-horizon overtone outburst discussed in Ref.~\cite{Konoplya:2022pbc}. Its ADM-normalized fundamental frequency is already appreciably displaced from Schwarzschild. The real-part difference grows gradually through $n=5$ and then decreases slightly, while the relative damping difference decreases throughout the sequence. This is consistent with the visible modification of the main potential barrier and distinguishes the present parameter choice from a deformation localized almost exclusively next to the horizon.}

{Finally, the complex-contour calculation reproduces the published Schwarzschild excitation factors and approaches them continuously from within the deformed family. In the asymptotic tortoise-coordinate convention, the excitation factors change little for moderate deformation but can vary dramatically in the extreme sector, with the electromagnetic quadrupole magnitude enhanced by a factor of about $140$ in one example. We have shown that this variation is predominantly associated with the large coordinate offset between the potential maximum and the asymptotically defined origin. Aligning the potential maxima reduces the changes in the excitation-factor magnitudes to modest deviations from Schwarzschild. The excitation factors in either convention are not direct observable amplitudes, since the full response also depends on the normalized radial functions and source-dependent excitation integral. Taken together, the frequency and residue calculations show that the geometry preserves damped test-field ringing and that the physical interpretation of the spectral and excitation properties requires particular care when the near-horizon and asymptotic mass scales differ substantially.}

Another important characteristic of radiation phenomena around black holes that has not been investigated here is the grey-body factors of scattered waves \cite{Konoplya:2026qql}. The accurate quasinormal frequencies obtained in our work could be used in future studies to test the correspondence between quasinormal modes and grey-body factors \cite{Konoplya:2024lir,Konoplya:2024vuj,Bolokhov:2024otn,Malik:2024cgb,Han:2026fpn,Han:2025cal,Calza:2026etq}.
\vspace{7mm}

\begin{acknowledgments}
The authors thank Sergei V. Bolokhov for useful discussions. The OpenAI was used to edit the text and prepare this manuscript. The authors assume the full responsibility for the final form of the manuscript. 
\end{acknowledgments}

\sloppy
\bibliographystyle{apsrev4-1}
\bibliography{refsNEW}

\end{document}